\documentclass[runningheads]{llncs}

\usepackage{eccv}

\usepackage{eccvabbrv}
\usepackage{enumitem}
\usepackage{graphicx}
\usepackage{booktabs}
\usepackage[accsupp]{axessibility}

\usepackage{hyperref}
\hypersetup{
    colorlinks=true,
    urlcolor=magenta
}

\usepackage{orcidlink}
\usepackage{multirow}
\usepackage{amsmath}
\usepackage{amssymb}
\usepackage{pifont}
\usepackage{pdflscape}

\newcommand{\x}{\mathbf{x}}
\newcommand{\p}{\mathbf{p}}
\newcommand{\norm}[1]{\left\lVert#1\right\rVert}

\DeclareMathOperator{\argmin}{argmin}
\DeclareMathOperator{\CNN}{CNN}
\DeclareMathOperator{\NO}{NO}
\newcommand{\cmark}{\ding{51}}%
\newcommand{\xmark}{\ding{55}}%

\def\method{%
  \textsc{CTO}%
  \futurelet\next\checknext
}
\def\checknext{%
  \ifx\next,%
  \else\ifx\next.%
  \else\ifx\next'%
  \else\ifx\next)%
  \else
    \ %
  \fi\fi\fi\fi
}

\title{Resolution-Agnostic Neural Operators for Multi-Rate Sparse-View CT}

\titlerunning{Resolution-Agnostic Neural Operators for Sparse-View CT}

\DeclareRobustCommand{\authorblock}{%
\setlength{\tabcolsep}{5pt}%
\begin{tabular}{@{}cccc@{}}
Aujasvit Datta$^{1,3\dagger}$\; \; \;  &
Jiayun Wang$^{1,2\dagger}$ \; \; &
Asad Aali$^{4}$  \; \; &
Anima Anandkumar$^{1}$
\\[3pt]
\multicolumn{4}{c}{$^1$Caltech \quad $^2$Georgia Tech \quad $^3$IIT Kanpur \quad $^4$Stanford University}
\\[3pt]
\multicolumn{4}{c}{\tt\footnotesize aujasvitd22@iitk.ac.in,\ \{peterw,\,anima\}@caltech.edu,\ asadaali@stanford.edu}
\\[3pt]
\multicolumn{4}{c}{$^{\dagger}$Equal contribution}
\end{tabular}%
}
\author{\authorblock}

\authorrunning{Datta et al.}

\institute{}
\begin{document}
\maketitle
\vspace{-1em}
\begin{abstract}
Sparse-view Computed Tomography (CT) reconstructs images from a limited number of X-ray projections to reduce radiation and scanning time, which is an ill-posed inverse problem. 
Existing methods achieve high-fidelity reconstructions but overfit to a fixed acquisition setup, failing to generalize well across sampling rates.
For example, convolutional neural networks (CNNs) use the same kernels across resolutions, leading to artifacts when data resolution changes. This is a critical limitation in clinical practice, where acquisition sampling settings vary across organs and diagnostic protocols.

We propose Computed Tomography neural Operator (CTO), the first neural operator (NO) framework for CT reconstruction. CTO extends learning from fixed discretized grids to continuous function space, enabling a single model to generalize across measurement sampling rates  without retraining. We also propose new NO architectural designs for CT: 
(i) a dual-domain NO architecture in both sinogram and image spaces, capturing complementary spatial–frequency information, and (ii) rotation-equivariant DIScrete–COntinuous convolutions (DISCO)  that exploit the rotational structure inherent in tomographic acquisition.
Empirically, CTO outperform CNNs ($>3.4$dB PSNR gain) and other baselines in multi-resolution settings across multiple CT datasets. Compared to state-of-the-art diffusion methods, CTO has $500\times$ faster inference with an average $3$dB gain.
CTO further demonstrates strong out-of-distribution robustness, maintaining gains under cross-dataset transfer and noisy sinogram conditions. Ablation studies also validate each design choice.
CTO establishes neural operators as a principled and practical paradigm for flexible, discretization-agnostic CT reconstruction. Our code is available at \url{https://github.com/neuraloperator/sparse_ct}.

\end{abstract}

 \vspace{-1.5em}
\section{Introduction}
\begin{figure*}[t]
  \centering
    \vspace{-1em}
  \includegraphics[width=\linewidth]{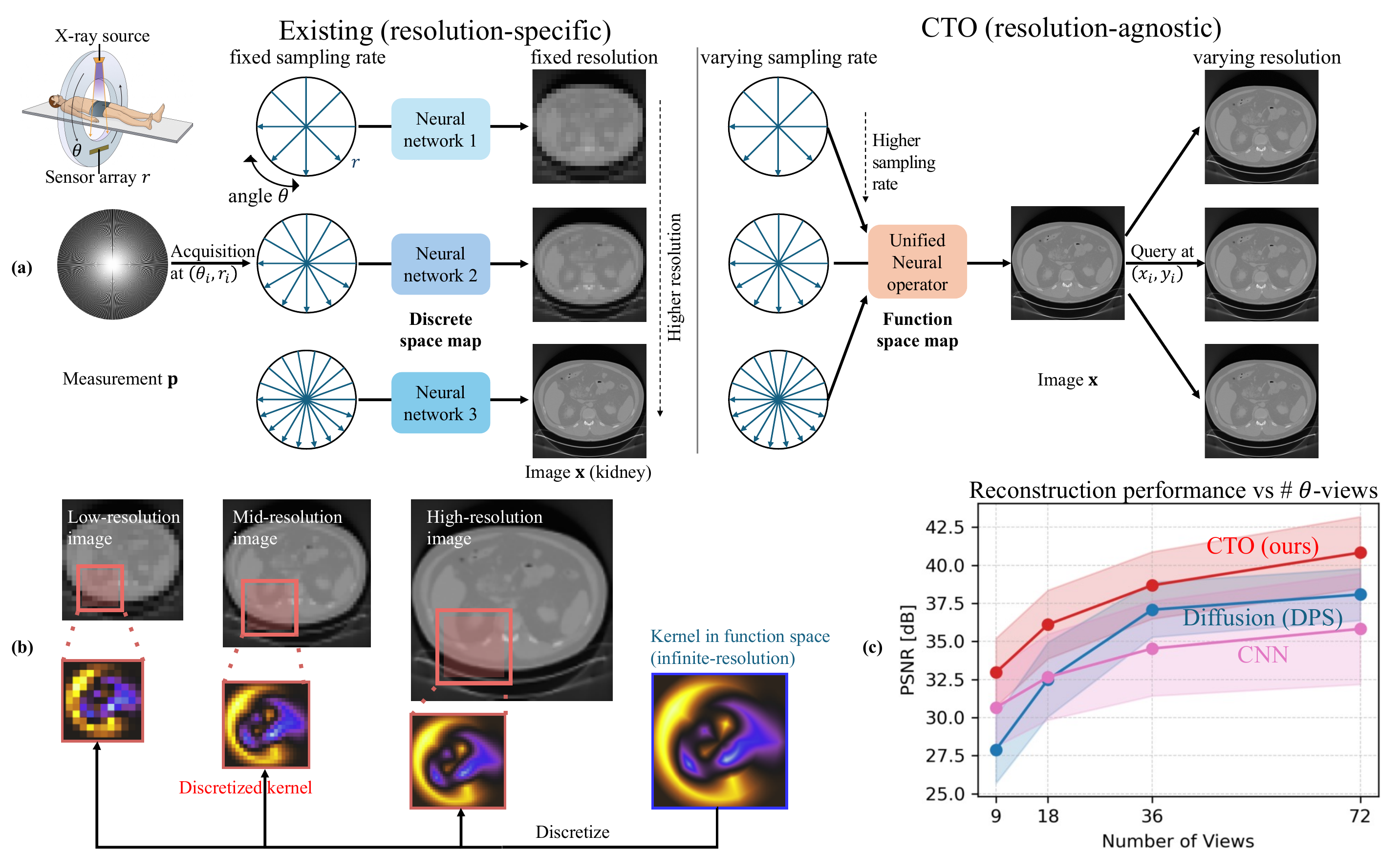}
  \caption{  {\bf (a)} We propose {\bf CTO}, a unified {\bf CT} reconstruction framework based on resolution-agnostic neural {\bf O}perator parameterized in function space. CTO reconstructs CT images across different measurement sampling rates and resolutions, whereas existing methods \cite{ayad2024qn,wang2022dudotrans} are resolution-dependent and need to be trained separately for different sampling rates due to resolution-specific backbones.
  {\bf (b) } CTO's basic building block is DISCO (discrete-continuous convolution) \cite{ocampo2022scalable}, a convolution layer defined in function space which works for images at different resolutions. Specifically, DISCO learns convolution kernels in the infinite-dimensional function space and can be discretized to a specific resolution based on the input data resolution to ensure a consistent receptive field across resolutions.
  {\bf (c)} CTO consistently outperforms the CNN baseline \cite{modl_19} across different sampling rates for kidney CT   \cite{heller2019kits19}. They both follow an unrolled network design with similar model size; CTO uses resolution-agnostic DISCO while CNN uses regular convolutions.}
  \label{fig:overview}
  \vspace{-1em}
\end{figure*}

Computed Tomography (CT) is a widely used imaging modality that reconstructs cross-sectional images of internal anatomy from X-ray projections acquired at multiple angles. The measured projection data (sinogram) represent line integrals of the object’s linear attenuation coefficient along different ray paths, mathematically modeled (for parallel-beam) by the Radon transform \cite{hermena2023ct, radon20051}. CT image reconstruction aims to recover the underlying attenuation map from these projections, forming an inverse problem—namely, inversion of the Radon transform. Under ideal conditions with fully sampled data and sufficient angular coverage, this inverse transform yields accurate reconstructions.
However, in sparse-view CT (SVCT), where the number of projection angles is significantly reduced and uniformly subsampled to lower radiation dose or shorten scan time \cite{han2018framing}, the inverse problem becomes ill-posed. In such cases, direct inversion leads to severe noisy artifacts, and additional regularization or learning-based methods are required to achieve high-quality reconstructions.

Classical CT reconstruction relies on filtered back projection (FBP) \cite{shepp1974fourier}, which is efficient but produces severe artifacts under sparse sampling \cite{abbas2013effects}.
Other classical compressed sensing methods \cite{candes2006robust,donoho2006compressed,sidky2008image}  improve quality by enforcing data consistency and sparsity priors, but at high computational cost. Deep learning has since transformed SVCT reconstruction: CNNs \cite{chen2017lowa,chen2017lowb} and transformer-based architectures \cite{zhang2021transct} substantially outperform classical methods. Unrolled model-based networks \cite{modl_19,adler2018learned,chen2018learn} bridge the gap between classical and learned approaches by unfolding iterative optimization into trainable network layers, embedding data consistency directly into the architecture. However, most of these approaches operate solely in the image domain, treating reconstruction as a post-processing step after FBP and discarding the rich information present in the raw sinogram. Dual-domain methods \cite{lin2019dudonet,wang2022dudotrans,zhou2022dudoufnet,wang2021indudonet} address this by jointly performing sinogram completion and image refinement with improved performance.

Yet, existing learning methods are tied to a specific discretization setting with a fixed input measurement sampling rate—a model trained for one measurement sampling rate does not generalize well to another \cite{liu2023dolce,jatyani2025unified} (See Table~\ref{tab:combined}a). A common solution to avoid sampling-rate-overfitting (e.g., \cite{shao2025multi}) is to train several models, each for a different sampling rate.
However, maintaining distinct models for different sparsity regimes limits scalability, and the approach does not generalize across sampling rates/image resolutions. This is a critical limitation in clinical settings, where acquisition settings like measurement sampling rates and target resolution constantly change across organs, protocols, and diagnostic purposes. {\it Thus, this paper adopts a setting where a unified model is used across acquisition sampling rates.}  This enables flexible CT reconstruction under varying conditions, thus broadening its clinical applicability.%

\vspace{-0.8em}
\paragraph{Our approach.}
We propose \method, a unified CT reconstruction framework based on resolution-agnostic Neural Operators (NOs)~\cite{kovachki2023neural} that learns mappings between infinite-dimensional function spaces, enabling a single model to work and generalize across measurement sampling rates and image resolutions without retraining.
To the best of our knowledge, \method is the first application of neural operators to sparse-view CT.%

\method uses unrolled networks with two U-shaped DISCO \cite{liu2024neural,jatyani2025unified} blocks that process the measurements and images: ($\NO_s$) in the {\it sinogram domain} and ($\NO_i$) in the {\it image domain}. The sinogram operator ($\NO_s$) is defined in polar coordinates $(r, \theta)$ — the natural space of CT measurements — and jointly learns in both {\it spatial and frequency spaces} of sinograms, inspired by FBP \cite{shepp1974fourier}.  The image operator ($\NO_i$), defined on the 2D Cartesian grid, further refines spatial features and improves reconstruction quality. Both operators are built on Discrete--Continuous (DISCO) convolutions~\cite{liu2024neural,jatyani2025unified}, which parametrize kernels in the continuous function space and discretize them to match different input or output resolution---making the entire framework inherently resolution-agnostic. DISCO has been proven to be discretization-convergent~\cite{berner2025principled,liu2024neural}: the approximation error is bounded across resolutions and converges to zero as resolution increases.

This design yields two structural properties. First, \method is resolution- and sampling-agnostic: the function-space parametrization allows consistent performance across different data resolutions without architectural changes. Second, learning in the sinogram domain's polar coordinate system makes \method rotation-equivariant---a rotation in the measurement coordinates induces an equivalent rotation in the reconstructed image---improving robustness to varying acquisition angles.
Empirically, replacing CNNs with the proposed neural operators yields an average gain of over $3.4$~dB PSNR across sinogram subsampling rates. \method outperforms state-of-the-art diffusion methods~\cite{chung2023dps} by over $3$~dB while being at least $500\times$ faster at inference. \method also outperforms all SOTA baselines in SVCT reconstruction and zero-shot super-resolution reconstruction.

Our main contributions are:
\textbf{1)} We propose \method, a unified end-to-end CT reconstruction framework based on neural operators that generalizes across measurement sampling rates and image resolutions. To the best of our knowledge, we are the first to use Neural Operators for sparse CT reconstruction.
\textbf{2)} We introduce a joint spatial-frequency sinogram-domain operator which respects the CT rotational geometry and acquisition physics.
\textbf{3)} By learning in the function space, \method is inherently rotation-equivariant and resolution-agnostic via DISCO operators in both sinogram and image domains, achieving consistent reconstructions across varying conditions. 
In summary, our central novelty is a {\it unified function-space formulation for sparse-view CT}: instead of learning separate discrete models for each sampling rate, we learn a single continuous operator from which different discretizations can be sampled, enabling {\it multi-rate reconstruction and cross-discretization consistency without retraining}.

The remainder of the paper is organized as follows. Sec.~\ref{sec:related_works} reviews related work on accelerated CT reconstruction and neural operators. Sec.~\ref{sec:ct_recon} establishes the theoretical framework for the problem, formulating CT reconstruction in function space and introducing DISCO convolutions. Sec.~\ref{sec: cto_design} details \method's design, including its dual-domain operators and rotation-equivariant convolution. Sec.~\ref{sec:expr} presents experimental results, and Sec.~\ref{summary} concludes.

\vspace{-1em}
\section{Related works}
\vspace{-0.5em}
\label{sec:related_works}

\noindent {\bf Accelerated CT Reconstruction}.
Methods broadly fall into (i) \textit{physics-based iterative} and (ii) \textit{learning-based} approaches. Classical compressed-sensing methods impose handcrafted priors, especially total variation (TV) and prior-image constrained CS (PICCS), to stabilize ill-posed sparse-view or limited-angle reconstruction. They reduce artifacts, but rely on slow multi-iteration solvers and degrade under extreme undersampling \cite{chen2008prior, liu2013total, yu2023reconstruction}, highlighting the \textit{speed--fidelity trade-off} of optimization-based CT.
Learning-based methods instead learn data-driven priors for faster inference \cite{ma2023freeseed, ayad2024qn, hu2023cross}. Early image-domain CNNs exploit hierarchical feature extraction and nonlinear representation capacity to outperform classical methods in sparse-view reconstruction \cite{chen2017lowa, chen2017lowb}, while architectures such as TransCT \cite{zhang2021transct} introduce dual-path transformers to refine high-frequency details through piecewise reconstruction. However, these methods treat reconstruction as post-hoc denoising of an initial FBP image, discarding rich sinogram information and often producing artifacts that violate physical consistency \cite{ye2018deep, zhang2021accurate}. Model-based unrolled networks such as Learned Primal-Dual (LPD) \cite{adler2018learned}, RegFormer \cite{xia2022transformer}, LEARN \cite{chen2018learn}, and QN-Mixer \cite{ayad2024qn} address this by integrating forward and adjoint operators for improved data consistency, but they remain tied to fixed sampling patterns, limiting adaptability. Dual-domain networks \cite{lin2019dudonet, zhou2022dudodr, zhou2022dudoufnet, wang2021indudonet, sun2024efficient, wang2022dudotrans, li2023learning, wang2021improving, wu2025dual} jointly perform sinogram completion and image refinement, addressing aliasing artifacts at their source and showing greater robustness.
Yet across these families, models remain bound to specific discretization settings: networks trained for one measurement sampling rate or output resolution generalize poorly to others \cite{liu2023dolce, jatyani2025unified}. Shao et al. \cite{shao2025multi} broaden coverage by synergistically training separate models for ultra-sparse and sparse regimes, but maintaining distinct networks for each regime limits scalability and does not generalize well across sampling rates. Fan et al.~\cite{fan2024mvmsrcn} train a single dual-domain unfolding model across multiple sparse-view rates, but remain CNN-based and tied to fixed discretizations, unable to generalize across image resolutions. Generative diffusion models \cite{zhou2025sparse, chen2025cross, song2024diffusionblend, liu2023dolce} achieve SOTA fidelity in SVCT by alternating data-consistency and denoising steps, but their high inference cost limits practical use. Across all families, the main challenges remain \cite{cheng5366340deep}: (1) discretization binding to specific sampling rates and resolutions and (2) inference-time scalability. {\it Our method addresses these challenges by introducing a NO formulation that learns in function space, enabling cross-resolution generalization and rotation-equivariant reconstruction without retraining.}

\noindent {\bf Neural Operators for Computational Imaging}.
For resolution-agnostic image reconstruction, 
diffusion models have shown empirically stable performance across data resolutions in accelerated CT reconstruction \cite{liu2023dolce,zhou2025sparse}. Yet, they typically incur higher inference times and require substantial hyperparameter tuning to achieve optimal results. Further, diffusion models are not inherently discretization-agnostic.
In contrast, a neural operator (NO) \cite{azizzadenesheli2024neural,kovachki2023neural} is inherently resolution-agnostic as it learns a map  between infinite-dimensional function spaces,
enabling evaluation at arbitrary resolutions and converging toward the target operator as the grid resolution increases. 
NOs were first used for accelerating the numerical solutions to partial differential equations (PDEs) \cite{kovachki2023neural,wang2024beyond,wang2025accelerating3dphotoacousticcomputed}. More recently, NOs have been used for resolution-agnostic computational imaging tasks in various domains, e.g., MRI \cite{jatyani2025unified}, neuro-imaging ~\cite{tolooshams2025vars}, and ultrasound imaging~\cite{wang2025ultrasound,wang2025accelerating3dphotoacousticcomputed}.
NO architectures are tailored to each application. For instance, the Fourier NO (FNO) \cite{fnop} employs global convolutions and has achieved robust discretization-agnostic performance across diverse tasks \cite{azizzadenesheli2024neural}. Other NOs \cite{gnop,liu2024neural} use locally-supported kernels to better capture localized structures, which is beneficial in applications involving localized dynamics such as turbulent fluid modeling \cite{gnop}. Additionally, NOs with local integral formulations are more amenable to parallel computation and can offer improved efficiency compared to architectures that rely on global operations.
{\it Our CT reconstruction framework leverages neural operators with local integral kernels, making it inherently agnostic to both sinogram sampling rates and output image resolution.}

\begin{figure*}[t]
  \centering
  \vspace{-1em}
  \includegraphics[width=0.9\linewidth]{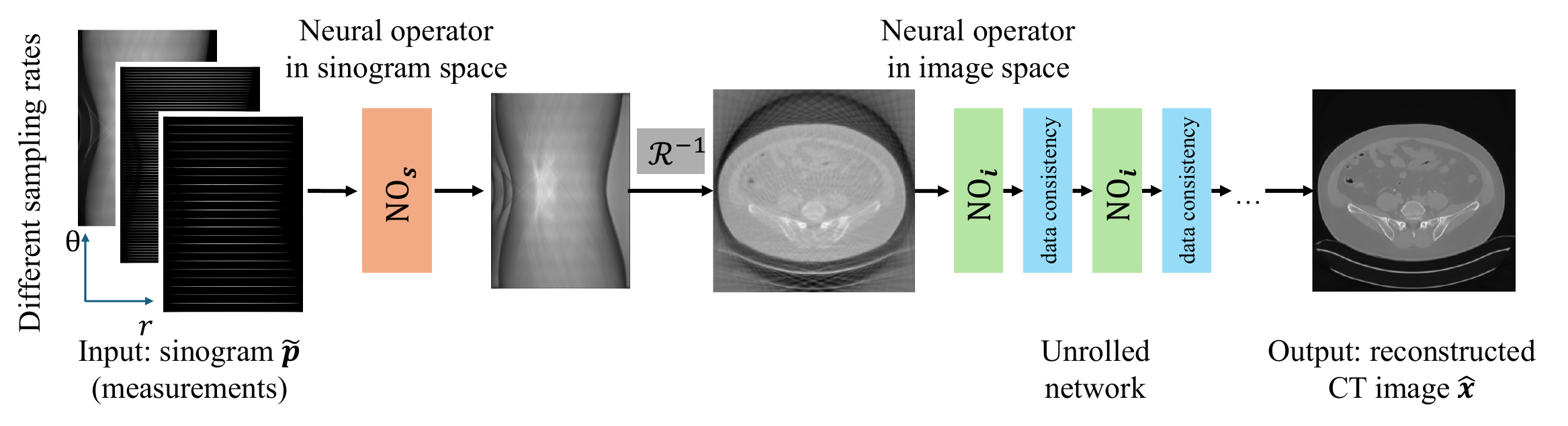}
  \vspace{-1em}
  \caption{
  {\bf \method architecture overview}. It follows an unrolled network design (Sec.~\ref{sec:unrll}). The input sensory signal sinogram is first fed to neural operator $\NO_s$ defined in sinogram space, allowing \method to be resolution-agnostic to different sampling rates for sinograms. Then, we use unrolled networks with multiple cascades mimicking a classical iterative algorithm, with each cascade consisting of image-space $\NO_i$ and a data consistency (physical update) term. They are both defined in function space and make \method discretization-agnostic to sampling rates and image resolutions.
  $\mathcal{R}^{-1}$ refers to the inverse Radon transform that transforms a sinogram to an image.}
  \label{fig:architect}
  \vspace{-1em}

\end{figure*}

\section{CT Reconstruction in Function Space}
\label{sec:ct_recon}
\method combines unrolled networks and function space learning to allow physical awareness (of the forward operator) and resolution-agnosticism  (Fig.~\ref{fig:architect}). Among multiple architectural designs \cite{fnop,gnop} available for neural operators, we choose discrete-continuous convolution (DISCO), especially for image reconstruction due to its similarity to convolutional layers. We finally present a U-shaped DISCO block, the basic building block enabling multi-scale receptive field.
\subsection{CT Reconstruction with Unrolled Networks} 
\label{sec:unrll}
In CT, cross-sectional images are reconstructed from a series of X-ray measurements acquired at multiple angles around an object (Fig.~\ref{fig:overview}a). As the beams propagate through the object, their intensity is attenuated according to the material along each path, and detectors record this attenuation as a one-dimensional projection. Collecting projections over many angles produces a two-dimensional sinogram, which compactly represents the measurement data.

Mathematically, the sinogram can be described as a function with polar coordinates $\p(\theta, r)$, where $\theta$ is the projection angle and $r$ is the detector position.  The sinogram's relationship with the underlying image $x$ can be written as
\begin{align}
    \p(\theta, r) := \mathcal{R}(\x) + \epsilon
\end{align}
where $\mathcal{R}$ is the Radon transform and $\epsilon$ is the measurement noise.
For a specific machine, sensor position $r$ is fixed, but the sampling of $\theta$ varies via the programming of the scanning speed.
To accelerate acquisition and reduce radiation, fewer projections are taken and measurements are subsampled as $\tilde{\p} = M\p$, where $M$ is an angle sampling operator, which can be implemented as a binary mask that selects a subset of projection angles. Recovering the image $\x$ from $\tilde{\p}$ becomes an ill-posed inverse problem.
Classical compressed sensing methods solve the inverse problem and reconstruct the underlying image $\hat{\x}$ via optimization  
\begin{align}
\label{eq:opt}
    \hat{\x} = \argmin_{\x} \frac{1}{2}  \norm{ \mathcal{A}( \x) - \tilde{\p}}_2^2 + \lambda \Psi(\x)
\end{align}
where $\mathcal{A}(\cdot) := M \mathcal{R}$ is the linear forward operator, $\Psi(\x)$ is a regularization term and $\lambda$ is the weight of the regularization term. The optimization can be solved with gradient descent. 
Recently,  unrolled networks \cite{adler2018learned, xia2022transformer, chen2018learn, modl_19} approximate the iterative solver  by cascading learned updates. We adopt \cite{modl_19}, a popular choice that uses multiple cascades of deep learning architectures such as CNNs, where a specific cascade $t$ mimics a gradient descent step from $\x^{t}$ to $\x^{t+1}$:
\begin{align}
\label{eq:cnn}
    \x^{t+1} &\leftarrow  \x^t-\eta^t\mathcal{A}^*(\mathcal{A}(\x^t)-\tilde{\p})+ \lambda^t \CNN^t(\x^t)%
\end{align}
where $\eta^t, \lambda^t$ are weights of the  data-consistency term and deep-learning-based regularization term. $\mathcal{A}^*$ is the hermitian of the
forward operator $\mathcal{A}$.

\noindent\textbf{Reconstruction in Function Space.}
Both sinograms (measurements) and images are \emph{functions}: $\p:\mathbb{S}^1\!\times\!\mathbb{R}\!\to\!\mathbb{R}$ and $\x:\Omega\!\subset\!\mathbb{R}^2\!\to\!\mathbb{R}$. A pixel grid is merely a discretization (sampling) of pixel value $\x$ over coordinates $(x,y)$; higher image resolution corresponds to denser $(x,y)$ sampling. Likewise, acquisition with more view angles corresponds to denser sampling of $\theta$ for $\p$.

While Eqn.~\eqref{eq:opt} naturally accommodates different samplings via $\mathcal{A}$, CNN-based unrolled networks (Eqn.~\eqref{eq:cnn}) optimize in a \emph{discrete subspace} tied to a fixed grid and often overfit to that resolution. In contrast, neural operators define the learned map directly between \emph{function spaces}, enabling a single model to act consistently across discretizations of both $\p$ and $\x$. We therefore propose to replace the discrete CNN prior with an image-space neural operator $\NO_i$:
\begin{align}
\label{eq:no}
    \x^{t+1} \;\leftarrow\; \x^t - \eta^t\,\mathcal{A}^*\big(\mathcal{A}(\x^t)-\tilde{\p}\big) + \lambda^t\,\NO^t_i(\x^t),
\end{align}
so the learned update is resolution-agnostic: the same model applies to different data discretizations (Fig.~\ref{fig:overview}a), i.e., different angle samplings for $\p$ and to different spatial resolutions for $\x$. We choose a neural operator design specifically adapted for computational imaging as follows.%

\subsection{DISCO (Discrete-Continuous) Convolutions: Basic Unit for Function Space Learning} 
\label{sec:disco}
The fundamental building block of the proposed neural operator is the discrete-continuous convolution (DISCO) neural operator \cite{liu2024neural}, which mimics the convolution layer but optimizes in the function space. It also provides a principled way to embed local inductive biases into neural operators while remaining resolution-agnostic. Classical CNNs achieve locality through finite-support convolutional kernels. However, a CNN layer converges to pointwise linear mappings as input resolution increases, thus losing its interpretation as local integral operators in the infinite-resolution limit \cite{liu2024neural}. This limits their ability to generalize across varying resolutions and downsampling patterns.
DISCO addresses this by defining the kernel as a continuous function in the function space and discretizing it only at the resolution of the input.
For a kernel $\kappa$ and input $g$ over a compact domain $D \subset \mathbb{R}^d$, the continuous convolution is given by
   $ (\kappa \star g)(v) = \int_D \kappa(u - v)\, g(u)\, du$
and is approximated by
\begin{align}
\vspace{-1em}
(\kappa \star g)(v_i) \approx \sum_{j=1}^m \kappa(u_j - v_i)\, g(u_j)\, q_j.
\end{align}
The kernel is parameterized as $\kappa = \sum_{\ell=1}^L \theta_\ell \kappa_\ell$ with learnable coefficients $\theta_\ell$ and fixed basis functions $\kappa_\ell$, where $\kappa_\ell$ is a piecewise-linear basis following \cite{jatyani2025unified} (visualized in Fig.~\ref{fig:overview}b). Because the basis is defined in the function space, the convolution scales with the domain rather than the grid, converging to a local integral operator as resolution increases. This property makes DISCO an effective backbone for both $\NO_s$ and $\NO_i$, enabling consistent feature learning across resolutions and sampling conditions.

\noindent \textbf{UDNO (U-Shaped DISCO Block).} UDNO \cite{jatyani2025unified} is a core architectural unit. Inspired by U-Net \cite{ronneberger2015u}, it progressively downsamples to aggregate coarse contextual information and upsamples to recover fine spatial detail. U-shaped networks \cite{ronneberger2015u} have shown strong performance across a range of vision tasks \cite{ronneberger2015u, croitoru2023diffusion, peebles2023scalable} due to their ability to learn multi-scale features. UDNO builds on this design but replaces standard convolutions with DISCO layers, combining multi-scale feature learning with the resolution-agnostic properties of NOs.

\section{CTO Design Details}
\label{sec: cto_design}

\begin{figure*}[t]
  \centering
    \vspace{-1em}
  \includegraphics[width=0.95\linewidth]{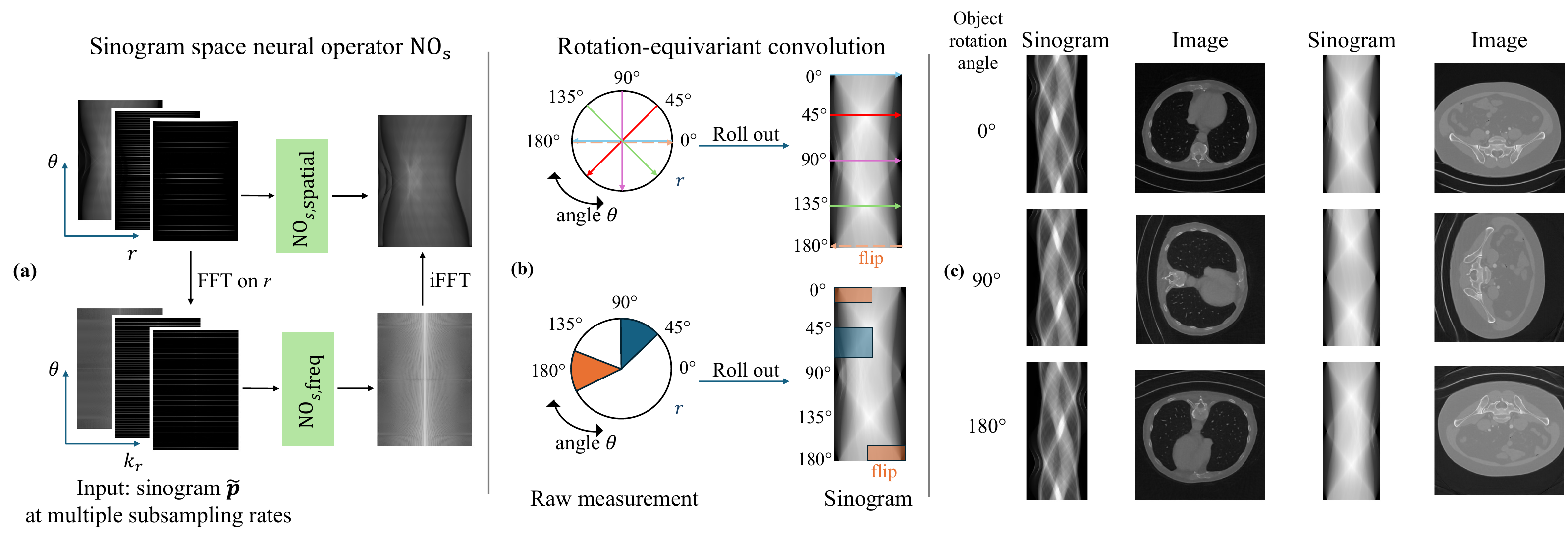}
      \vspace{-1em}
    \caption{
(a) \textbf{Sinogram-space neural operator $\NO_s$.} The subsampled sinogram $\tilde{\p}$ is processed by two complementary operators: $\NO_{s,\text{spatial}}$ in the spatial domain and $\NO_{s,\text{freq}}$ in the frequency domain via a 1D Fourier transform along the detector axis $r$. Their outputs are combined and passed to the image-space stage.
(b) \textbf{Rotation-equivariant learning.} Object rotation corresponds to a shift along the angular axis $\theta$ in the sinogram. Circular padding along $\theta$ preserves this rotational structure during learning. Note the angle flips at $180^\circ$ (See Supp.~Sec.~\ref{sec:roteq}).
(c) \textbf{Visualization of rotational equivariance.} We rotate the object by $0^\circ$, $90^\circ$, and $180^\circ$ and show the corresponding sinograms (left) and \method reconstructions (right). Reconstructions rotate consistently with the inputs, confirming stable performance across orientations.
    }
      \label{fig:rot_eq_visualization}
    \vspace{-1em}
\end{figure*}

\textbf{Framework Overview.}
The proposed \method is an end-to-end model following the unrolled network architecture \footnote{We use a cascaded dual-domain design because it is necessary for strong state-of-the-art CT performance, but our main contribution lies in replacing discretization-specific operators with a single function-space operator that supports multiple view samplings and image resolutions within one model.}(Fig.~\ref{fig:architect}).
Specifically, the input sinogram $\p$ is first fed to a sinogram space neural operator $\NO_s$ for function space learning that unifies different sampling rates for $\theta$. The intermediate feature map is then fed to the unrolled network for multi-cascade learning, where the image-space neural operator $\NO_i$ and the physical update/data-consistency term denoise and reconstruct the final images (Eqn.~\eqref{eq:no}).

DISCO in $\NO_i$ is defined in 2D Cartesian coordinates, introducing a local inductive bias in the image space that captures fine details across discretizations. In $\NO_s$, we extend DISCO from Cartesian to polar coordinates $(r, \theta)$, allowing the kernels to provide localized support in the sinogram domain. Due to Radon transform duality, localized support corresponds to direction-selective global support in the image space, thus allowing $\NO_s$ to learn global structural information. The two components jointly model complementary information about the image, resulting in improved structural consistency and sharper detail.

\subsection{Sinogram-Space Operator $\NO_s$}
\label{sec:rotational_equivalence}
\textbf{Sinogram-Space NO}. 
We introduce a sinogram-space neural operator $\NO_s$ that acts on the function space of $\p(\theta,r)$, with detector position $r$ and view angle $\theta$ treated as the two sinogram axes. DISCO convolutions are applied in this $(\theta,r)$ coordinate system—without projection or reparameterization to $(x,y)$—so the model learns directly in the intrinsic measurement domain, avoiding projection errors. Extending DISCO from 2D Cartesian to polar coordinates requires honoring the periodic boundary condition in $\theta$; we therefore use circular padding along $\theta$ (and standard padding along $r$), yielding $\theta$-shift equivariance (rotational equivariance) and consistent handling of arbitrary angular offsets. As a neural \emph{operator}, $\NO_s$ maps between function spaces rather than fixed grids, so the same model applies across discretizations (e.g., different detector bins).
 
{We adapt DISCO, originally defined in the image space, to the sinogram space with a novel dual-branch design.} 
The $\NO_s$ module comprises two parallel UDNO branches (Fig.~\ref{fig:rot_eq_visualization}a): $\NO_{s,\text{spatial}}$ and $\NO_{s,\text{freq}}$. $\NO_{s,\text{spatial}}$ operates on $\p(\theta,r)$ to capture local, geometry-aware correlations, while $\NO_{s,\text{freq}}$ operates on the sinogram’s 2D frequency representation to model long-range angular–radial dependencies. The two outputs are averaged to form the final sinogram-space estimation, which is then forwarded to the downstream reconstruction stage. The design is inspired by FBP \cite{shepp1974fourier}, where the additional $r$-frequency processing of the sinogram greatly reduces reconstruction blur.

\noindent \textbf{Rotation-Equivariant Convolution}.
Parallel-beam CT induces two symmetries on the sinogram $\p(\theta,r)$:
(i) rotation of the image by $\phi$ is a \emph{shift in view angle} of the sinogram,
$(\p \!\circ\! R_\phi)(\theta,r) \;=\; \p(\theta-\phi,r),$
and (ii) 
$
\p(\theta+\pi,r)\;=\;\p(\theta,-r)
$ (the $\pi$-periodicity-with-flip identity). 
We design $\NO_s$ to be rotation-equivariant (see Supp.~\ref{sec:roteq}). For implementation, $\NO_s$ adopts a circular padding along the projection angle ($\theta$) axis of the sinogram to account for the inherent periodicity of projection data. Projections at $\theta = 0$ and $\theta = 180^\circ$ correspond to parallel beam paths and are therefore identical up to a flip along the detector $r$ axis. More generally, the projection at $\theta = 180^\circ + \alpha$ is equivalent to the projection at $\theta = \alpha$ after a flip along the detector $r$ axis. To preserve this physical continuity, we use a modified circular padding scheme that first flips the boundary data along the detector $r$ axis and then wraps it around the projection angle $\theta$ axis.

We visualize the equivariant design in Fig.~\ref{fig:rot_eq_visualization}b.
This design choice is significant for two reasons. 1) it provides physically consistent boundary conditions in $\NO_s$, preventing discontinuities at $\theta = 0$ and $\theta = 180^\circ$ and the artifacts they can cause. 2) It enhances rotational robustness by aligning the network’s inductive biases with the geometry of the sinogram domain. Since DISCO operators in $\NO_s$ act over the $(r, \theta)$ axes, which we treat as rectangular axes for NO learning, rotational transformations of the image $x$ manifest as translations along these axes, which can be naturally modeled by the translational equivariance of DISCO. As this periodicity arises only along the angular dimension, we apply flipped circular padding along $\theta$ and standard reflective padding along $r$.

\noindent \textbf{UDNO for Spatial Sinogram.} $\NO_{s,\text{spatial}}$ is a UDNO with circular padding, which learns in the spatial domain of the sinogram. Mathematically, its output $\hat{\p}_{\text{spatial}}$ can be written as
$
\hat{\p}_{\text{spatial}} = \NO_{s, \text{spatial}} (\tilde{\p})
$.

\noindent \textbf{UDNO for Sinogram's Frequency Component.} To learn in the frequency domain of the sinogram, we first apply a one-dimensional Fourier transform along the detector ($r$) axis of the sinogram. The transformed representation is then processed by the circular-padded UDNO $\NO_{s, \text{freq}}$, after which an inverse Fourier transform is applied along the same axis to return the output to the spatial domain. The overall output $\hat{\p}_{\text{freq}}$ can be written as
$
\hat{\p}_{\text{freq}} = \mathcal{F}_r^{-1}(\NO_{s, \text{freq}}(\mathcal{F}_{r}(\tilde{\p}(r, \theta))))
$.

This design is motivated by two complementary considerations. First, the Fourier slice theorem provides a theoretical foundation for learning in this domain. The theorem states that the one-dimensional Fourier transform of the projection along $r$ satisfies
\vspace{-0.5em}
\begin{align}
    \mathcal{F}_r\{\p(\theta, r)\}(\omega) = \hat{f}(\omega\cos\theta, \omega\sin\theta),
\end{align}
where $\hat{f}(\xi_x, \xi_y)$ is the two-dimensional Fourier transform of the object. This result implies that the Fourier transform of the sinogram encodes the object’s frequency spectrum in polar coordinates. Consequently, localized operations in this domain correspond to global modifications in the reconstructed image, allowing $\NO_{s,\text{freq}}$ to learn high-level global features about the underlying image.

Second, classical reconstruction algorithms such as FBP incorporate an explicit prior on the frequency domain of the sinogram by applying fixed frequency-domain filters (such as the ramp filter) to the one-dimensional Fourier transform of the sinogram along the detector ($r$) axis, suppressing low-frequency components before backprojection. Our approach can be interpreted as a data-driven generalization of this idea: rather than imposing a hand-crafted prior, $\NO_{s, \text{freq}}$ learns a frequency-domain prior directly from data. This learned prior captures how the sinogram should be completed under different subsampling conditions, thus reducing image artifacts and enabling the model to generalize better.

\noindent \textbf{Final Output. } The final output of $\NO_s$ is given by:
$
\NO_s(\tilde{\p}) = \frac{\hat{\p}_{\text{spatial}} + \hat{\p}_{\text{freq}}}{2}
$,
which is then passed downstream for processing in the image domain.
\subsection{Image-Space Operator $\NO_i$}
\label{sec:NOI}
$\NO_i$ is implemented as a UDNO (Sec.~\ref{sec:disco})   that operates directly in the spatial domain of the image space. DISCO allows $\NO_i$ to train and infer at different image resolutions. 
$\NO_i$ refines and denoises the intermediate reconstruction obtained from the sinogram-space operator by capturing high-frequency local information and finer structural details that may not be fully recovered in earlier stages. This ensures final reconstructions preserve multi-scale features, including large-scale structural consistency and small-scale details.

\section{Experimental Results}
\label{sec:expr}
We study a setting in which a single model operates across varying sinogram sampling rates (Fig.~\ref{fig:overview}a-Right). Using sampling rate as an example, we observe that existing approaches~\cite{ayad2024qn,chen2018learn}  train separate models for each rate. Such rate-specific training overfits to the corresponding configuration (Table~\ref{tab:combined}a, row 1-2). In contrast, training on multiple-sampling-rate data mitigates this issue and improves generalization (Table~\ref{tab:combined}a, row 3-5). For fairness, all methods and baselines are trained using the same amount of multi-rate data. Baseline method hyperparameters are tuned to ensure the best multi-rate performance. 

\subsection{Dataset and Setup}
\label{sec:dataset}
\noindent{\bf Datasets}:
We consider two CT datasets in the paper:
\begin{itemize}
    \item \textbf{AAPM Low-Dose Abdominal CT Dataset}  \cite{mccollough2016tu} consists of 5{,}936 axial slices with patient ID.
    Following the split in \cite{wang2022dudotrans, li2023learning}, we select 90\% of patients for training and the rest (526 slices) for testing. 
    \item \textbf{Kidney CT Scan Dataset} \cite{heller2019kits19} is from 2019 Kidney and Kidney Tumor Segmentation Challenge (C4KC-KiTS). For each CT volume, we select the central 50\% of slices and a uniform 10\% sample of peripheral slices to ensure coverage across anatomical variability, resulting in 33{,}000 training slices from 170 patients. For evaluation, we use 10{,}000 slices from 40 held-out patients.

\end{itemize}
Main-text quantitative tables focus on AAPM \cite{mccollough2016tu}; Kidney \cite{heller2019kits19} visualization and results are discussed in main text and further expanded in Supp.~\ref{sec:kidney}.
\begin{table*}[t]
\centering
\setlength{\tabcolsep}{5pt}
\renewcommand{\arraystretch}{1.2}
\caption{Sparse-view CT reconstruction results for AAPM
dataset \cite{mccollough2016tu}. Lower is better for RMSE; higher is better for PSNR/SSIM. For all learning methods, a single model is trained on multi-view sinograms with $N_v = {9, 18, 36, 72}$ views together. For fairness, all methods are trained using the same amount of data -- this differs from the resolution-specific models trained in their original paper since multi-resolution training generally reduces overfitting (Table~\ref{tab:combined}a). All methods are reported on the entire test set. 9-view test results are in Supp.~\ref{sec:9view} due to space limit.}
\label{tab:aapm_multires_results}
\resizebox{\textwidth}{!}{%
\begin{tabular}{llccccccccc}
\toprule
\multirow{2}{*}{{\bf Category}} & \multirow{2}{*}{{\bf Method}} &
\multicolumn{3}{c}{{\bf 18-view}} & \multicolumn{3}{c}{{\bf 36-view}} & \multicolumn{3}{c}{{\bf 72-view}} \\
\cmidrule(lr){3-5}\cmidrule(lr){6-8}\cmidrule(lr){9-11}
& & RMSE (HU)$\downarrow$ & PSNR$\uparrow$ & SSIM (x$10^{-2}$) $\uparrow$
  & RMSE (HU)$\downarrow$ & PSNR$\uparrow$ & SSIM (x$10^{-2}$) $\uparrow$
  & RMSE (HU)$\downarrow$ & PSNR$\uparrow$ & SSIM (x$10^{-2}$) $\uparrow$ \\
\midrule
\multirow{2}{*}{\begin{tabular}{c}Learning-free\end{tabular}}
& FBP \cite{shepp1974fourier}    & 632.58 $\pm$ 42.17 & 13.14 $\pm$ 1.63 & 33.87 $\pm$ 6.32 & 616.66 $\pm$ 41.07 & 13.36 $\pm$ 1.63 & 36.07 $\pm$ 5.89 & 584.72 $\pm$ 38.87 & 13.82 $\pm$ 1.63 & 40.35 $\pm$ 5.66 \\
& SART \cite{andersen1984simultaneous}  & 161.91 $\pm$ 5.93  & 24.97 $\pm$ 1.46 & 59.51 $\pm$ 4.53 & 145.15 $\pm$ 2.59 & 25.91 $\pm$ 1.41 & 69.09 $\pm$ 3.37 & 133.24 $\pm$ 1.81  & 26.66 $\pm$ 1.40 & 76.85 $\pm$ 2.06 \\
\midrule
\multirow{2}{*}{\begin{tabular}{c}Diffusion\end{tabular}}
& DPS \cite{chung2023dps}           & 149.97 $\pm$ 17.34 & 25.68 $\pm$ 1.77 & 60.08 $\pm$ 8.28 & 100.50 $\pm$ 7.33 & 29.13 $\pm$ 1.51 & 76.71 $\pm$ 4.02 & 75.19 $\pm$ 4.71 & 31.64 $\pm$ 1.43 & 82.61 $\pm$ 2.94 \\

& ALD \cite{jalal2021robust}         & 148.13 $\pm$ 14.63  & 25.78 $\pm$ 1.71 & 63.42 $\pm$ 7.19 & 110.19 $\pm$ 8.78 & 28.33 $\pm$ 1.46 & 75.50 $\pm$ 3.92 & 85.17 $\pm$ 5.35  & 30.56 $\pm$ 1.41 & 80.47 $\pm$ 3.12 \\
\midrule
\multirow{3}{*}{\begin{tabular}{c}Single pass\end{tabular}}
& DuDoTrans \cite{wang2022dudotrans} & 154.64 $\pm$ 6.18 & 25.36 $\pm$ 1.50 & 69.52 $\pm$ 4.32 & 92.17 $\pm$ 3.59 & 29.88 $\pm$ 1.62 & 78.21 $\pm$ 3.59 & 69.28 $\pm$ 4.99 & 32.35 $\pm$ 1.57 & 81.49 $\pm$ 3.38\\
& GloReDi \cite{li2023learning}        & 96.89 $\pm$ 10.39  & 29.47 $\pm$ 1.70 & 78.52 $\pm$ 3.69 & 74.42 $\pm$ 6.24 & 31.74 $\pm$ 1.64 & 83.59 $\pm$ 3.05 & 64.63 $\pm$ 4.34  & 32.96 $\pm$ 1.55 & 85.78 $\pm$ 2.60 \\
& MDPRNet \cite{shao2025multi}        & 151.78 $\pm$ 17.64  & 25.38 $\pm$ 1.82 & 73.05 $\pm$ 4.53 & 116.85 $\pm$ 11.28 & 27.91 $\pm$ 1.79 & 81.60 $\pm$ 3.85 & 74.02 $\pm$ 8.93  & 30.87 $\pm$ 1.55 & 87.18 $\pm$ 3.64 \\
\midrule
\multirow{5}{*}{\begin{tabular}{c}Unrolled Networks\end{tabular}}
& Learned PD \cite{adler2018learned} & 155.14 $\pm$ 13.54  & 25.35 $\pm$ 1.61 & 50.21 $\pm$ 5.62 & 126.36 $\pm$ 9.79 & 27.13 $\pm$ 1.60 & 57.7 $\pm$ 5.82 & 91.07 $\pm$ 5.70  & 29.96 $\pm$ 1.53 & 71.88 $\pm$ 4.59 \\
& LEARN \cite{chen2018learn} & 153.39 $\pm$ 22.22  & 25.51 $\pm$ 1.94 & 73.42 $\pm$ 3.90 & 121.31 $\pm$ 16.15 & 27.53 $\pm$ 1.81 & 77.64 $\pm$ 3.41 & 92.49 $\pm$ 12.91  & 29.90 $\pm$ 1.59 & 81.37 $\pm$ 2.72 \\
& RegFormer \cite{xia2022transformer} & 150.57 $\pm$ 21.34  & 25.67 $\pm$ 1.91 & 72.95 $\pm$ 4.02 & 114.93 $\pm$ 15.18 & 28.00 $\pm$ 1.82 & 77.87 $\pm$ 3.56 & 86.83 $\pm$ 10.46  & 30.43 $\pm$ 1.70 & 81.55 $\pm$ 3.01 \\
& Unrolled CNN \cite{modl_19} & 100.65 $\pm$ 11.36  & 29.14 $\pm$ 1.84 & 78.21 $\pm$ 5.47 & 74.44 $\pm$ 8.32 & 31.76 $\pm$ 1.80 & 82.01 $\pm$ 4.95 & 62.38 $\pm$ 11.16  & 33.35 $\pm$ 1.81 & 84.49 $\pm$ 5.27 \\
& \textbf{CTO (ours)} & \textbf{75.97 $\pm$ 7.32} & \textbf{31.57 $\pm$ 1.73} & \textbf{81.62 $\pm$ 4.02} & \textbf{50.79 $\pm$ 4.27} & \textbf{35.06 $\pm$ 1.64} & \textbf{87.14 $\pm$ 3.12} & \textbf{36.64 $\pm$ 2.44} & \textbf{37.88 $\pm$ 1.56} & \textbf{91.55 $\pm$ 1.81} \\
\bottomrule
\end{tabular}
}
\end{table*}

\noindent \textbf{\method Implementation.} \method follows  unrolled networks \cite{modl_19} with 3 cascades. Each cascade consists of an image-space operator ($
\NO_i$) followed by a data consistency module. All UDNOs (Sec.~\ref{sec:disco}) share the same architectural configuration: DISCO parameterized with a piecewise-linear kernel basis \cite{jatyani2025unified} composed of one isotropic basis and 5 anisotropic basis rings, each containing 7 basis functions (hyperparameter-tuned, CNN baseline matches the model size). Each UDNO uses one input and one output channel, 32 hidden channels, and four encoder–decoder (pooling) levels. The DISCO kernels use a radius cutoff of 0.02 and are defined over an input domain of size $256 \times 256$. Models are trained using mean squared error loss with the Adam optimizer and a learning rate of $1e-3$.

\noindent \textbf{Baseline: Unrolled Networks.} We compare with existing SOTA Model-based Unrolled Networks like Learned Primal-Dual \cite{adler2018learned}, RegFormer \cite{xia2022transformer}, and LEARN \cite{chen2018learn}. We also compare with an Unrolled CNN model which uses the same architecture as \method, with DISCOs replaced by regular convolutions. Unrolled CNN can be considered an ablation study that compares neural operator (function space) and neural network (discrete space) architecture. However, Unrolled CNN has slightly different number of parameters than \method due to a different kernel configuration.

\noindent \textbf{Baseline: Single-pass methods.} We compare with single-pass methods like DuDoTrans\cite{wang2022dudotrans}, GloReDi\cite{li2023learning} and MDPRNet \cite{shao2025multi}.
Details of baseline implementations, sinogram sampling settings, hardware and training are in Supp.~Sec.~\ref{sec:implement_details}.

\noindent \textbf{Evaluation Protocols.} CT images represent linear attenuation coefficients that are typically interpreted in \emph{Hounsfield Units} (HU), where water is defined as 0 HU and air as $-1000$ HU. Following \cite{chen2025cross}, we report RMSE after converting reconstructed images to HU to provide a clinically meaningful measure of reconstruction error. We also report PSNR and SSIM in the original reconstruction space (before HU conversion) to assess structural and perceptual fidelity independent of scanner calibration. Thus, RMSE (HU) reflects quantitative clinical accuracy, while PSNR and SSIM capture image similarity. 
\vspace{-1em}
\subsection{Reconstruction with Different Sampling Rates}
\label{sec:details}

\begin{figure*}[t]
  \centering
 \vspace{-1em}
  \includegraphics[width=0.95\linewidth]{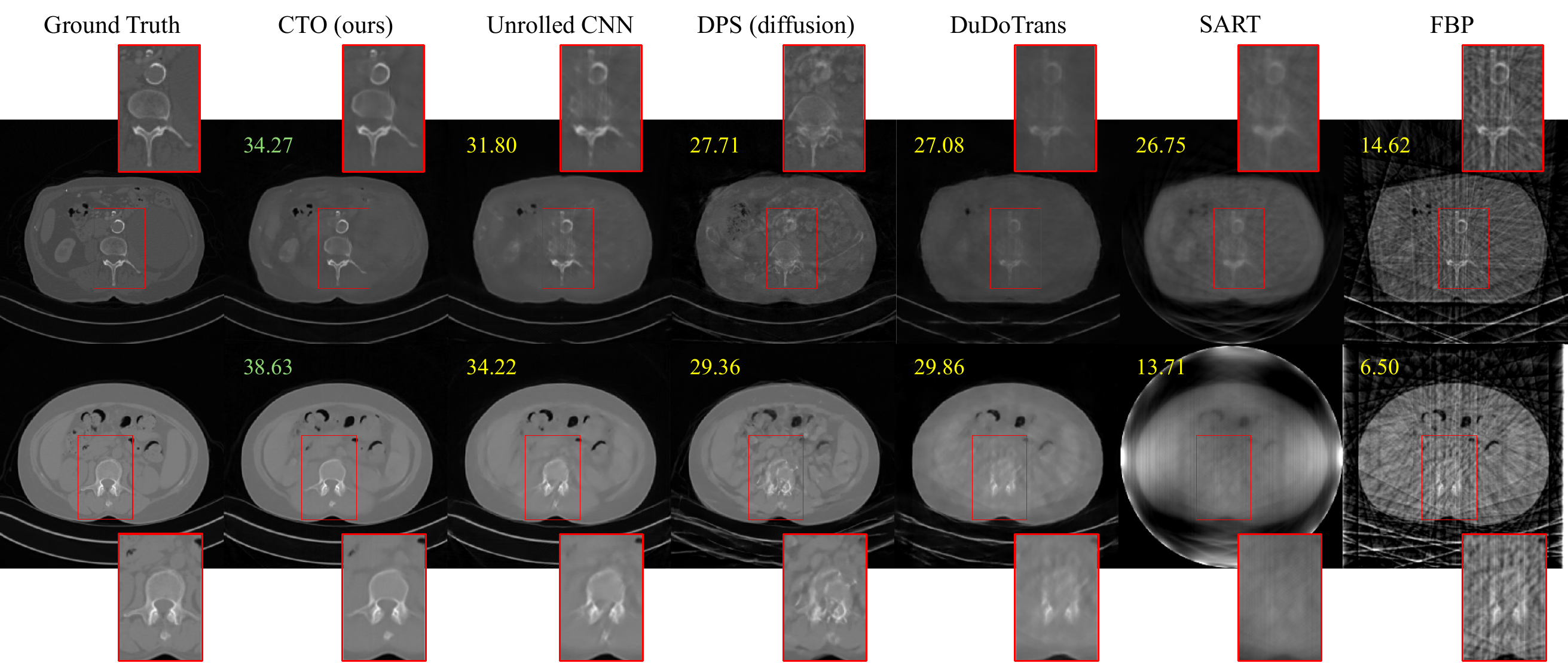}
      \vspace{-0.5em}
    \caption{
Reconstruction on 18-view sampling rate for Abdomen Low-Dose CT dataset \cite{mccollough2016tu} (top row) and Kidney dataset \cite{heller2019kits19} (bottom row). \method outperforms baselines for PSNR (upper left) and reconstruction quality.
    }
      \label{fig:visual}
    \vspace{-1em}
\end{figure*}

\noindent \textbf{Multi-Sparse-View Rate Results}. We train and evaluate reconstruction performance at 9-, 18-, 36-, and 72-view acquisition settings, where fewer projections correspond to lower radiation at the cost of more ill-posed reconstruction. For all methods, a single model is trained across all sampling rates. In each training batch, a sparsity rate is randomly and uniformly chosen, ensuring that all models train on equal amounts of data for each sparsity rate. Results for the AAPM dataset \cite{mccollough2016tu} are shown in Table~\ref{tab:aapm_multires_results}, and for Kidney dataset in Fig.~\ref{fig:overview}c.
Full results for the C4KC-KiTS Kidney dataset \cite{heller2019kits19} are reported in Supp.~\ref{sec:kidney}. 

\method outperforms learning-free methods \cite{shepp1974fourier,andersen1984simultaneous} and deep learning methods, including both diffusion methods \cite{chung2023dps,jalal2021robust} and architectures like CNNs \cite{modl_19} and transformers \cite{wang2022dudotrans}.
\noindent On the kidney dataset \cite{heller2019kits19}, on average across different numbers of views, we perform 1) 4 dB PSNR better than unrolled CNN baseline; 2) 3 dB PSNR better than diffusion-DPS \cite{chung2023dps}; 3) 5 dB better than transformer-DuDoTrans \cite{wang2022dudotrans}.
\noindent On the abdomen dataset \cite{mccollough2016tu}, we perform 1) 3 dB PSNR better than unrolled CNN baseline; 2) 6 dB better than DPS \cite{chung2023dps}; 3) 6 dB better than   DuDoTrans \cite{wang2022dudotrans} on average. Visualizations are in Fig.~\ref{fig:visual}. Note that some baseline models show different performance from what was reported in their original papers. This is expected since they train multiple models for each rate.

\begin{figure}[t] 
    \vspace{-1em}
    \centering
    \includegraphics[width=0.8\linewidth]{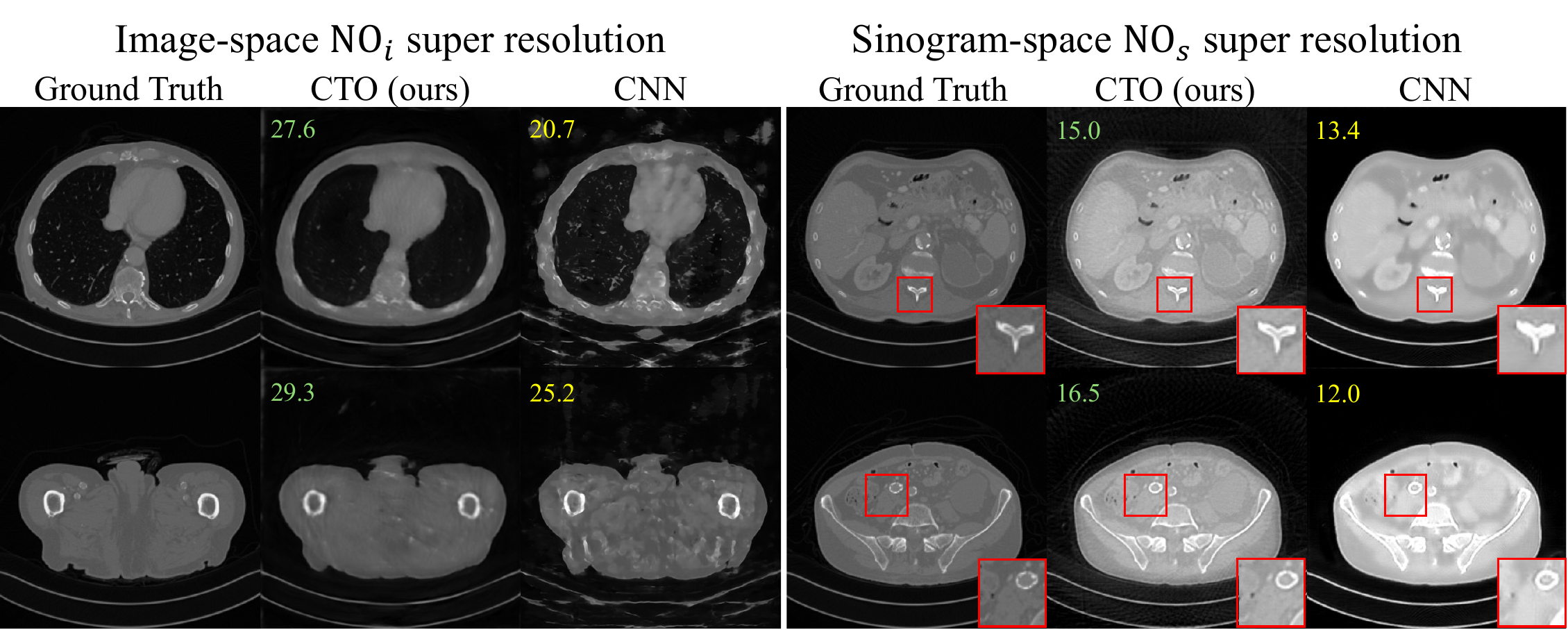}
    
    \caption{Zero-shot super resolution results. \method outperforms CNN baseline for both $\NO_i$ and $\NO_s$ super resolution.}
    \label{fig:superres}
    \vspace{-1em}
\end{figure}

\subsection{Zero-Shot Super-Resolution Results}

\noindent \textbf{Super-Resolution on $\NO_i$.} We evaluate the zero-shot super-resolution capability of NO in the image domain. All models—\method, the Unrolled CNN baseline \cite{modl_19}, RegFormer \cite{xia2022transformer}, and LEARN \cite{chen2018learn}—are trained to produce $256 \times 256$ reconstructions from $18$-view sinograms. At inference, we test generalization to a higher resolution \emph{without any fine-tuning}: the sinogram-space operator $\NO_s$ is kept fixed, and the intermediate $256 \times 256$ output is bilinearly upsampled to $512 \times 512$ before image-space processing. Reconstructions are evaluated against fully sampled ground truth images that have been bilinearly interpolated to $512\times512$. When the resolution doubles, CNN's fixed-size kernels cover half the relative receptive field, whereas a NO's kernel, defined as a continuous function, maintains the same relative coverage. Consequently, \method achieves a 3 dB PSNR gain over the CNN baseline, and 4.8 dB gain over LEARN \cite{chen2018learn}. As shown in Fig.~\ref{fig:superres}, the CNN baseline exhibits substantial blurring and aliasing at the higher resolution, while \method{} preserves much sharper detail.

\noindent \textbf{Super-Resolution on $\NO_s$.} We evaluate zero-shot super-resolution in the sinogram domain by training all models—\method, the Unrolled CNN baseline \cite{modl_19}, RegFormer \cite{xia2022transformer}, and LEARN \cite{chen2018learn}—at one sinogram sampling setting and testing at another. Models are trained on sinograms with $72$ views and $672$ detectors, then evaluated on two increased-sinogram-size settings: 1) increased number of acquisition views: sinograms with $144$ views and $672$ detectors (\method outperforms Unrolled CNN by $3.5$ dB PSNR, LEARN by $3.2$ dB PSNR, RegFormer by $2.8$ dB PSNR); 2) further increased number of detectors: sinograms with $144$ views and $1344$ detectors (\method outperforms Unrolled CNN by $3.7$ dB PSNR, LEARN by $2.7$ dB PSNR, RegFormer by $2.4$ dB PSNR.).
As shown in Fig.~\ref{fig:superres}, \method{} preserves fine anatomical structures with fewer artifacts compared to the baselines.

\vspace{-2.7mm}

\subsection{Analysis and Ablation Study}
\label{sec:ablation}

\begin{table*}[t]
\vspace{-0.5em}
\centering
\caption{{\bf (a) Left:} Baseline methods overfit to the seen sampling rate, performing substantially worse on other rates (row 1-2). Training on all rates together (row 3-5) mitigates this problem. We adopt multi-rate co-training for all methods for the all-view setting (row 3-5). {\bf (b) Right:} Inference and tuning time of methods tested on NVIDIA A100. \method is $>500\times$ faster than diffusion \cite{jalal2021robust}.}
\label{tab:combined}
\vspace{-0.5em}
\begin{minipage}[t]{0.45\linewidth}
\centering
\resizebox{\linewidth}{!}{
\begin{minipage}[t]{0.5\linewidth}
\centering
\resizebox{\linewidth}{!}{
\begin{tabular}{lccc}
\toprule
Method/training rate & \multicolumn{3}{c}{Test rate} \\
& {\bf 18-view} & {\bf 36-view} & {\bf 72-view} \\
\midrule
LEARN-72 view      &  4.16     &    22.40   &    32.32   \\
RegFormer-72 view  &   5.04    &   24.63    &    35.11   \\
\midrule
LEARN-all view     &    25.51   &   27.53    &   29.90    \\
RegFormer-all view & 25.67 & 28.0  & 30.43        \\
CTO-all view       & {\bf 31.57} & {\bf 35.06} & {\bf 37.88}        \\
\bottomrule
\end{tabular}}
\end{minipage}
}
\end{minipage}%
\hfill
\begin{minipage}[t]{0.54\linewidth}
\centering
\resizebox{\linewidth}{!}{
\begin{tabular}{llcc}
\toprule
{\bf Category} & \textbf{Method} & \textbf{Inference Time (s)} & \textbf{Tuning Required} \\
\midrule
Learning-free & SART \cite{andersen1984simultaneous} & 0.417 & \cmark \\ \midrule
Diffusion & DPS \cite{chung2023dps} & 51.58 & \cmark \\
& ALD \cite{jalal2021robust} & 32.72 & \cmark \\ \midrule
Single-pass & RegFormer \cite{xia2022transformer} & 0.085 \\
 & {\method (ours)} & {\bf 0.065} & \xmark \\
\bottomrule
\end{tabular}%
}
\end{minipage}
\vspace{-1em}
\end{table*}

\noindent \textbf{Rotational Equivariance.} As discussed in Sec.~\ref{sec:rotational_equivalence}, circular padding enforces the periodic boundary conditions inherent to the sinogram's angular domain, enabling the network to better exploit rotational equivariance. Supp.~Table~\ref{tab:circular_padding} shows that this design leads to a $0.83$ dB PSNR gain under $24$-view subsampling (33 dB with circular padding, 32.17 dB without circular padding). Qualitative results (Fig.~\ref{fig:rot_eq_visualization}c) further demonstrate that the reconstruction rotates as the object rotates, confirming that the model captures true rotational structure. Our equivariant design gives a notable improvement in PSNR, indicating that handling angular periodicity explicitly enhances reconstruction quality.

\noindent \textbf{Ablation of $\NO_s$ and $\NO_i$.} Averaged over view counts, removing $\NO_i$ and $\NO_s$ individually leads to PSNR drops of 3.86 dB and 4.82 dB, respectively. Within $\NO_s$, removing the spatial branch causes a 3.1 dB drop and the frequency branch a 2.7 dB drop. Further $\NO_s$ design ablations are in Supp.~Sec.~\ref{sec:supp_nos}.

\noindent \textbf{DISCO Kernel Hyperparameter Sensitivity.} We perform a Bayesian search over the kernel basis, kernel radius, number of basis rings, and bases per ring for DISCO in \method. The results are reported in Supp.~Sec.~\ref{sec:DISCOhyper}.

\noindent{\bf Model Inference and Tuning Time}. We compare the model development and inference time of our end-to-end NO with other methods (Table~\ref{tab:combined}b): Diffusion models are at least $500\times$ slower in inference and require pattern-specific hyperparameter tuning for each sampling rate, since each rate corresponds to a different measurement distribution and the same model cannot be applied directly across rates. Classical learning-free methods like SART \cite{andersen1984simultaneous} also require hyperparameter tuning for specific $\p$ sampling rates during optimization.

\noindent\textbf{Out of Distribution (OOD) Generalization.} We perform several experiments to test OOD generalization of \method. Results are in Supp.~Sec.~\ref{sec:suppOOD}.

\noindent\textbf{LIDC-IDRI Lung CT Dataset.} We also evaluate Unrolled CNN and \method on the LIDC-IDRI lung CT dataset \cite{armato2011lung} in Supp.~Sec.~\ref{sec:lidc}.

\noindent\textbf{Complexity analysis.} Comparisons with other NO architectures are in Supp. Sec.~\ref{sec:comp_NOs}, and with the Unrolled CNN baseline in Supp.~Sec.~\ref{sec:runtimecnn}.%

\section{Summary}
\label{summary}
We present \method, a unified neural operator-based framework for sparse-view CT reconstruction that generalizes across different sampling rates of sensory data, 
overcoming the limitations of existing deep learning methods, which are tied to a fixed acquisition sampling rate. Specifically, we adopt the Neural Operator, a deep learning framework that learns maps between infinite-dimensional function spaces and thus is agnostic to different data resolutions.
By combining rotation-equivariant discrete-continuous convolutions  (DISCO) (a specific neural operator design that closely mimics convolutions but uses function-space kernels) with joint spatial–frequency learning in the sinogram domain, \method achieves robust, resolution-agnostic reconstructions and consistently outperforms state-of-the-art methods across varying downsampling regimes, enabling a scalable and flexible alternative to traditional CT reconstruction techniques. Future work could extend \method to more clinically relevant datasets, 3D CT and other computational imaging tasks, conduct uncertainty analysis on \method, and evaluate its clinical applicability across broader scanner protocols. 

\section*{Acknowledgment}
This work is supported in part by ONR (MURI grant N000142312654 and N000142012786). A.D. is supported in part by the Undergraduate Research Fellowships (SURF) at Caltech. J.W. is supported in part by the Pritzker AI+Science initiative and Schmidt Sciences.   A.A. is supported in part by Bren endowed chair and the AI2050 senior fellow program at Schmidt Sciences.

\bibliographystyle{splncs04}
\bibliography{main}

@article{candes2006robust,
	title        = {Robust uncertainty principles: Exact signal reconstruction from highly incomplete frequency information},
	author       = {Cand{\`e}s, Emmanuel J and Romberg, Justin and Tao, Terence},
	year         = 2006,
	journal      = {IEEE Transactions on information theory},
	publisher    = {IEEE},
	volume       = 52,
	number       = 2,
	pages        = {489--509},
	doi          = {10.1109/TIT.2005.862083}
}

@article{donoho2006compressed,
	title        = {Compressed sensing},
	author       = {Donoho, David L},
	year         = 2006,
	journal      = {IEEE Transactions on information theory},
	publisher    = {IEEE},
	volume       = 52,
	number       = 4,
	pages        = {1289--1306},
	doi          = {10.1109/TIT.2006.871582}
}

@article{sidky2008image,
	title        = {Image reconstruction in circular cone-beam computed tomography by constrained, total-variation minimization},
	author       = {Sidky, Emil Y and Pan, Xiaochuan},
	year         = 2008,
	journal      = {Physics in Medicine \& Biology},
	publisher    = {IOP Publishing},
	volume       = 53,
	number       = 17,
	pages        = 4777,
	doi          = {10.1088/0031-9155/53/17/021}
}

@article{chen2008prior,
	title        = {Prior image constrained compressed sensing (PICCS): a method to accurately reconstruct dynamic CT images from highly undersampled projection data sets},
	author       = {Chen, Guang-Hong and Tang, Jie and Leng, Shuai},
	year         = 2008,
	journal      = {Medical physics},
	publisher    = {Wiley Online Library},
	volume       = 35,
	number       = 2,
	pages        = {660--663},
	doi          = {10.1118/1.2836423}
}

@inproceedings{jatyani2025unified,
	title        = {A unified model for compressed sensing mri across undersampling patterns},
	author       = {Jatyani, Armeet Singh and Wang, Jiayun and Chandrashekar, Aditi and Wu, Zihui and Liu-Schiaffini, Miguel and Tolooshams, Bahareh and Anandkumar, Anima},
	year         = 2025,
	booktitle    = {Proceedings of the Computer Vision and Pattern Recognition Conference},
	pages        = {26004--26013},
	doi          = {10.48550/arXiv.2410.16290}
}

@misc{wang2025accelerating3dphotoacousticcomputed,
	title        = {Accelerating 3D Photoacoustic Computed Tomography with End-to-End Physics-Aware Neural Operators},
	author       = {Jiayun Wang and Yousuf Aborahama and Arya Khokhar and Yang Zhang and Chuwei Wang and Karteekeya Sastry and Julius Berner and Yilin Luo and Boris Bonev and Zongyi Li and Kamyar Azizzadenesheli and Lihong V. Wang and Anima Anandkumar},
	year         = 2025,
	url          = {https://arxiv.org/abs/2509.09894},
	eprint       = {2509.09894},
	archiveprefix = {arXiv},
	primaryclass = {eess.IV},
	doi          = {10.48550/arXiv.2509.09894}
}

@article{wang2025ultrasound,
	title        = {Ultrasound lung aeration map via physics-aware neural operators},
	author       = {Wang, Jiayun and Ostras, Oleksii and Sode, Masashi and Tolooshams, Bahareh and Li, Zongyi and Azizzadenesheli, Kamyar and Pinton, Gianmarco F and Anandkumar, Anima},
	year         = 2025,
	journal      = {ArXiv},
	pages        = {arXiv--2501},
	doi          = {10.48550/arXiv.2501.01157}
}

@article{tolooshams2025vars,
	title        = {VARS-fUSI: Variable Sampling for Fast and Efficient Functional Ultrasound Imaging using Neural Operators},
	author       = {Tolooshams, Bahareh and Lin, Lydia and Callier, Thierri and Wang, Jiayun and Pal, Sanvi and Chandrashekar, Aditi and Rabut, Claire and Li, Zongyi and Blagden, Chase and Norman, Sumner L and others},
	year         = 2025,
	journal      = {bioRxiv},
	publisher    = {Cold Spring Harbor Laboratory},
	pages        = {2025--04},
	doi          = {10.1101/2025.04.16.649237}
}

@misc{wang2024beyond,
	title        = {Coarse Graining with Neural Operators for Simulating Chaotic Systems},
	author       = {Chuwei Wang and Julius Berner and Zongyi Li and Di Zhou and Jiayun Wang and Jane Bae and Anima Anandkumar},
	year         = 2025,
	url          = {https://arxiv.org/abs/2408.05177},
	eprint       = {2408.05177},
	archiveprefix = {arXiv},
	primaryclass = {cs.LG},
	doi          = {10.48550/arXiv.2408.05177}
}

@article{han2018framing,
	title        = {Framing U-Net via deep convolutional framelets: Application to sparse-view CT},
	author       = {Han, Yoseob and Ye, Jong Chul},
	year         = 2018,
	journal      = {IEEE transactions on medical imaging},
	publisher    = {IEEE},
	volume       = 37,
	number       = 6,
	pages        = {1418--1429},
	doi          = {10.1109/TMI.2018.2823768}
}

@article{hermena2023ct,
	title        = {CT-scan image production procedures},
	author       = {Hermena, Shady and Young, Michael},
	year         = 2023,
	journal      = {StatPearls}
}

@inproceedings{ronneberger2015u,
	title        = {U-net: Convolutional networks for biomedical image segmentation},
	author       = {Ronneberger, Olaf and Fischer, Philipp and Brox, Thomas},
	year         = 2015,
	booktitle    = {International Conference on Medical image computing and computer-assisted intervention},
	pages        = {234--241},
	organization = {Springer},
	doi          = {10.1007/978-3-319-24574-4_28}
}

@article{heller2019kits19,
	title        = {The kits19 challenge data: 300 kidney tumor cases with clinical context, ct semantic segmentations, and surgical outcomes},
	author       = {Heller, Nicholas and Sathianathen, Niranjan and Kalapara, Arveen and Walczak, Edward and Moore, Keenan and Kaluzniak, Heather and Rosenberg, Joel and Blake, Paul and Rengel, Zachary and Oestreich, Makinna and others},
	year         = 2019,
	journal      = {arXiv preprint arXiv:1904.00445},
	doi          = {10.48550/arXiv.1904.00445}
}

@article{shepp1974fourier,
	title        = {The Fourier reconstruction of a head section},
	author       = {Shepp, Lawrence A and Logan, Benjamin F},
	year         = 1974,
	journal      = {IEEE Transactions on nuclear science},
	publisher    = {IEEE},
	volume       = 21,
	number       = 3,
	pages        = {21--43},
	doi          = {10.1109/TNS.1974.6499235}
}

@article{liu2024neural,
	title        = {Neural operators with localized integral and differential kernels},
	author       = {Liu-Schiaffini, Miguel and Berner, Julius and Bonev, Boris and Kurth, Thorsten and Azizzadenesheli, Kamyar and Anandkumar, Anima},
	year         = 2024,
	journal      = {arXiv preprint arXiv:2402.16845},
	doi          = {10.48550/arXiv.2402.16845}
}

@article{modl_19,
   title={MoDL: Model-Based Deep Learning Architecture for Inverse Problems},
   volume={38},
   ISSN={1558-254X},
   url={http://dx.doi.org/10.1109/TMI.2018.2865356},
   DOI={10.1109/tmi.2018.2865356},
   number={2},
   journal={IEEE Transactions on Medical Imaging},
   publisher={Institute of Electrical and Electronics Engineers (IEEE)},
   author={Aggarwal, Hemant K. and Mani, Merry P. and Jacob, Mathews},
   year={2019},
   month=feb, pages={394–405} }

@article{zhou2025sparse,
	title        = {Sparse-view CT image reconstruction using conditional embedding fusion diffusion model},
	author       = {Zhou, Chenchun and Sun, Yubao and Liang, Jing and Liu, Jia and Liu, Qingshan},
	year         = 2025,
	journal      = {Neurocomputing},
	publisher    = {Elsevier},
	pages        = 131748,
	doi          = {10.1016/j.neucom.2025.131748}
}

@inproceedings{liu2023dolce,
	title        = {Dolce: A model-based probabilistic diffusion framework for limited-angle ct reconstruction},
	author       = {Liu, Jiaming and Anirudh, Rushil and Thiagarajan, Jayaraman J and He, Stewart and Mohan, K Aditya and Kamilov, Ulugbek S and Kim, Hyojin},
	year         = 2023,
	booktitle    = {Proceedings of the IEEE/CVF international conference on computer vision},
	pages        = {10498--10508},
	doi          = {10.1109/ICCV51070.2023.00963}
}

@article{radon20051,
	title        = {1.1 {\"u}ber die bestimmung von funktionen durch ihre integralwerte l{\"a}ngs gewisser mannigfaltigkeiten},
	author       = {Radon, Johann},
	year         = 2005,
	journal      = {Classic papers in modern diagnostic radiology},
	publisher    = {Springer Berlin--Heidelberg--New York},
	volume       = 5,
	number       = 21,
	pages        = 124
}

@article{abbas2013effects,
	title        = {Effects of sparse sampling schemes on image quality in low-dose CT},
	author       = {Abbas, Sajid and Lee, Taewon and Shin, Sukyoung and Lee, Rena and Cho, Seungryong},
	year         = 2013,
	journal      = {Medical physics},
	publisher    = {Wiley Online Library},
	volume       = 40,
	number       = 11,
	pages        = 111915,
	doi          = {10.1118/1.4825096}
}

@article{andersen1984simultaneous,
	title        = {Simultaneous algebraic reconstruction technique (SART): a superior implementation of the ART algorithm},
	author       = {Andersen, Anders H and Kak, Avinash C},
	year         = 1984,
	journal      = {Ultrasonic imaging},
	publisher    = {Elsevier},
	volume       = 6,
	number       = 1,
	pages        = {81--94},
	doi          = {10.1177/016173468400600107}
}

@article{berner2025principled,
  title={Principled Approaches for Extending Neural Architectures to Function Spaces for Operator Learning},
  author={Berner, Julius and Liu-Schiaffini, Miguel and Kossaifi, Jean and Duruisseaux, Valentin and Bonev, Boris and Azizzadenesheli, Kamyar and Anandkumar, Anima},
  journal={arXiv preprint arXiv:2506.10973},
  year={2025},
  doi={10.48550/arXiv.2506.10973}
}

@article{adler2018learned,
	title        = {Learned primal-dual reconstruction},
	author       = {Adler, Jonas and {\"O}ktem, Ozan},
	year         = 2018,
	journal      = {IEEE transactions on medical imaging},
	publisher    = {IEEE},
	volume       = 37,
	number       = 6,
	pages        = {1322--1332},
	doi          = {10.1109/TMI.2018.2799231}
}

@article{kovachki2023neural,
	title        = {Neural operator: Learning maps between function spaces with applications to pdes},
	author       = {Kovachki, Nikola and Li, Zongyi and Liu, Burigede and Azizzadenesheli, Kamyar and Bhattacharya, Kaushik and Stuart, Andrew and Anandkumar, Anima},
	year         = 2023,
	journal      = {Journal of Machine Learning Research},
	volume       = 24,
	number       = 89,
	pages        = {1--97},
}

@article{croitoru2023diffusion,
	title        = {Diffusion models in vision: A survey},
	author       = {Croitoru, Florinel-Alin and Hondru, Vlad and Ionescu, Radu Tudor and Shah, Mubarak},
	year         = 2023,
	journal      = {IEEE transactions on pattern analysis and machine intelligence},
	publisher    = {Ieee},
	volume       = 45,
	number       = 9,
	pages        = {10850--10869},
	doi          = {10.1109/TPAMI.2023.3261988}
}

@inproceedings{peebles2023scalable,
	title        = {Scalable diffusion models with transformers},
	author       = {Peebles, William and Xie, Saining},
	year         = 2023,
	booktitle    = {Proceedings of the IEEE/CVF international conference on computer vision},
	pages        = {4195--4205},
	doi          = {10.1109/ICCV51070.2023.00387}
}

@article{gnop,
	title        = {Neural Operator: Graph Kernel Network for Partial Differential Equations},
	author       = {Zongyi Li and Nikola Kovachki and Kamyar Azizzadenesheli and Burigede Liu and Kaushik Bhattacharya and Andrew Stuart and Anima Anandkumar},
	year         = 2020,
	journal      = {arXiv preprint arXiv:2003.03485},
	eprint       = {2003.03485},
	archiveprefix = {arXiv},
	primaryclass = {cs.LG},
	doi          = {10.48550/arXiv.2003.03485}
}

@article{azizzadenesheli2024neural,
	title        = {Neural operators for accelerating scientific simulations and design},
	author       = {Azizzadenesheli, Kamyar and Kovachki, Nikola and Li, Zongyi and Liu-Schiaffini, Miguel and Kossaifi, Jean and Anandkumar, Anima},
	year         = 2024,
	journal      = {Nature Reviews Physics},
	publisher    = {Nature Publishing Group UK London},
	pages        = {1--9},
	doi          = {10.1038/s42254-024-00712-5}
}

@article{fnop,
	title        = {Fourier Neural Operator for Parametric Partial Differential Equations},
	author       = {Zongyi Li and Nikola Kovachki and Kamyar Azizzadenesheli and Burigede Liu and Kaushik Bhattacharya and Andrew Stuart and Anima Anandkumar},
	year         = 2021,
	journal      = {arXiv preprint arXiv:2010.08895},
	eprint       = {2010.08895},
	archiveprefix = {arXiv},
	primaryclass = {cs.LG},
	doi          = {10.48550/arXiv.2010.08895}
}

@article{mccollough2016tu,
  title={Overview of the low dose CT grand challenge},
  author={McCollough, Cynthia},
  journal={Medical physics},
  volume={43},
  number={6Part35},
  pages={3759--3760},
  year={2016},
  publisher={Wiley Online Library},
  doi={10.1118/1.4957556}
}

@article{liu2013total,
  title={Total variation-stokes strategy for sparse-view X-ray CT image reconstruction},
  author={Liu, Yan and Liang, Zhengrong and Ma, Jianhua and Lu, Hongbing and Wang, Ke and Zhang, Hao and Moore, William},
  journal={IEEE transactions on medical imaging},
  volume={33},
  number={3},
  pages={749--763},
  year={2013},
  publisher={IEEE},
  doi={10.1109/TMI.2013.2295738}
}

@article{yu2023reconstruction,
  title={Reconstruction of sparse-view x-ray computed tomography based on adaptive total variation minimization},
  author={Yu, Zhengshan and Wen, Xingya and Yang, Yan},
  journal={Micromachines},
  volume={14},
  number={12},
  pages={2245},
  year={2023},
  publisher={MDPI},
  doi={10.3390/mi14122245}
}

@inproceedings{ye2018deep,
  title={Deep back projection for sparse-view CT reconstruction},
  author={Ye, Dong Hye and Buzzard, Gregery T and Ruby, Max and Bouman, Charles A},
  booktitle={2018 ieee global conference on signal and information processing (globalsip)},
  pages={1--5},
  year={2018},
  organization={IEEE},
  doi={10.1109/GlobalSIP.2018.8646669}
}

@article{zhang2021accurate,
  title={Accurate and robust sparse-view angle CT image reconstruction using deep learning and prior image constrained compressed sensing (DL-PICCS)},
  author={Zhang, Chengzhu and Li, Yinsheng and Chen, Guang-Hong},
  journal={Medical physics},
  volume={48},
  number={10},
  pages={5765--5781},
  year={2021},
  publisher={Wiley Online Library},
  doi={10.1002/mp.15183}
}

@inproceedings{lin2019dudonet,
  title={DuDoNet: Dual domain network for CT metal artifact reduction},
  author={Lin, Wei-An and Liao, Haofu and Peng, Cheng and Sun, Xiaohang and Zhang, Jingdan and Luo, Jiebo and Chellappa, Rama and Zhou, Shaohua Kevin},
  booktitle={Proceedings of the IEEE/CVF Conference on Computer Vision and Pattern Recognition},
  pages={10512--10521},
  year={2019},
  doi={10.1109/CVPR.2019.01076}
}

@article{zhou2022dudodr,
  title={DuDoDR-Net: Dual-domain data consistent recurrent network for simultaneous sparse view and metal artifact reduction in computed tomography},
  author={Zhou, Bo and Chen, Xiongchao and Zhou, S Kevin and Duncan, James S and Liu, Chi},
  journal={Medical Image Analysis},
  volume={75},
  pages={102289},
  year={2022},
  publisher={Elsevier},
  doi={10.1016/j.media.2021.102289}
}

@article{zhou2022dudoufnet,
  title={DuDoUFNet: Dual-domain under-to-fully-complete progressive restoration network for simultaneous metal artifact reduction and low-dose CT reconstruction},
  author={Zhou, Bo and Chen, Xiongchao and Xie, Huidong and Zhou, S Kevin and Duncan, James S and Liu, Chi},
  journal={IEEE transactions on medical imaging},
  volume={41},
  number={12},
  pages={3587--3599},
  year={2022},
  publisher={IEEE},
  doi={10.1109/TMI.2022.3189759}
}

@inproceedings{wang2021indudonet,
  title={InDuDoNet: An interpretable dual domain network for CT metal artifact reduction},
  author={Wang, Hong and Li, Yuexiang and Zhang, Haimiao and Chen, Jiawei and Ma, Kai and Meng, Deyu and Zheng, Yefeng},
  booktitle={International Conference on Medical Image Computing and Computer-Assisted Intervention},
  pages={107--118},
  year={2021},
  organization={Springer},
  doi={10.1007/978-3-030-87231-1_11}
}

@inproceedings{wang2022dudotrans,
  title={DuDoTrans: dual-domain transformer for sparse-view CT reconstruction},
  author={Wang, Ce and Shang, Kun and Zhang, Haimiao and Li, Qian and Zhou, S Kevin},
  booktitle={International workshop on machine learning for medical image reconstruction},
  pages={84--94},
  year={2022},
  organization={Springer},
  doi={10.1007/978-3-031-17247-2_9}
}

@article{sun2024efficient,
  title={An efficient dual-domain deep learning network for sparse-view CT reconstruction},
  author={Sun, Chang and Salimi, Yazdan and Angeliki, Neroladaki and Boudabbous, Sana and Zaidi, Habib},
  journal={Computer methods and programs in biomedicine},
  volume={256},
  pages={108376},
  year={2024},
  publisher={Elsevier},
  doi={10.1016/j.cmpb.2024.108376}
}

@inproceedings{li2023learning,
  title={Learning to distill global representation for sparse-view ct},
  author={Li, Zilong and Ma, Chenglong and Chen, Jie and Zhang, Junping and Shan, Hongming},
  booktitle={Proceedings of the IEEE/CVF international conference on computer vision},
  pages={21196--21207},
  year={2023},
  doi={10.48550/arXiv.2308.08463}
}

@article{wu2025dual,
  title={Dual-Domain deep prior guided sparse-view CT reconstruction with multi-scale fusion attention},
  author={Wu, Jia and Lin, Jinzhao and Jiang, Xiaoming and Zheng, Wei and Zhong, Lisha and Pang, Yu and Meng, Hongying and Li, Zhangyong},
  journal={Scientific Reports},
  volume={15},
  number={1},
  pages={16894},
  year={2025},
  publisher={Nature Publishing Group UK London},
  doi={10.1038/s41598-025-02133-5}
}

@article{song2024diffusionblend,
  title={Diffusionblend: Learning 3d image prior through position-aware diffusion score blending for 3d computed tomography reconstruction},
  author={Song, Bowen and Hu, Jason and Luo, Zhaoxu and Fessler, Jeffrey and Shen, Liyue},
  journal={Advances in Neural Information Processing Systems},
  volume={37},
  pages={89584--89611},
  year={2024},
  doi={10.48550/arXiv.2406.10211}
}

@inproceedings{chen2025cross,
  title={Cross-View Generalized Diffusion Model for Sparse-View CT Reconstruction},
  author={Chen, Jixiang and Lin, Yiqun and Qin, Yi and Wang, Hualiang and Li, Xiaomeng},
  booktitle={International Conference on Medical Image Computing and Computer-Assisted Intervention},
  pages={140--150},
  year={2025},
  organization={Springer},
  doi={10.48550/arXiv.2508.10313}
}

@article{cheng5366340deep,
  title={Deep Learning for Sparse-View Ct Reconstruction: A Survey},
  author={Cheng, Shuaiqi and Chen, Yuxi and Yang, Bo and Liu, Chao and Zhang, Lijuan and Yin, Lirong and Zheng, Wenfeng},
  journal={Available at SSRN 5366340},
  year={2025}
}

@article{song2020score,
  title={Score-based generative modeling through stochastic differential equations},
  author={Song, Yang and Sohl-Dickstein, Jascha and Kingma, Diederik P and Kumar, Abhishek and Ermon, Stefano and Poole, Ben},
  journal={arXiv preprint arXiv:2011.13456},
  year={2020},
  doi={10.48550/arXiv.2011.13456}
}

@article{karras2022elucidating,
  title={Elucidating the design space of diffusion-based generative models},
  author={Karras, Tero and Aittala, Miika and Aila, Timo and Laine, Samuli},
  journal={Advances in neural information processing systems},
  volume={35},
  pages={26565--26577},
  year={2022},
  doi={10.48550/arXiv.2206.00364}
}

@inproceedings{chung2023dps,
  title={Diffusion Posterior Sampling for General Noisy Inverse Problems},
  author={Chung, Hyungjin and Kim, Jeongsol and Mccann, Michael Thompson and Klasky, Marc Louis and Ye, Jong Chul},
  booktitle={The Eleventh International Conference on Learning Representations},
  year={2023},
  doi={10.48550/arXiv.2209.14687}
}

@article{jalal2021robust,
  title={Robust compressed sensing mri with deep generative priors},
  author={Jalal, Ajil and Arvinte, Marius and Daras, Giannis and Price, Eric and Dimakis, Alexandros G and Tamir, Jon},
  journal={Advances in neural information processing systems},
  volume={34},
  pages={14938--14954},
  year={2021},
  doi={10.48550/arXiv.2108.01368}
}

@article{ocampo2022scalable,
  title={Scalable and equivariant spherical CNNs by discrete-continuous (DISCO) convolutions},
  author={Ocampo, Jeremy and Price, Matthew A and McEwen, Jason D},
  journal={arXiv preprint arXiv:2209.13603},
  year={2022},
  doi={10.48550/arXiv.2209.13603}
}

@inproceedings{ma2023freeseed,
  title={FreeSeed: Frequency-band-aware and self-guided network for sparse-view CT reconstruction},
  author={Ma, Chenglong and Li, Zilong and Zhang, Junping and Zhang, Yi and Shan, Hongming},
  booktitle={International conference on medical image computing and computer-assisted intervention},
  pages={250--259},
  year={2023},
  organization={Springer},
  doi={10.1007/978-3-031-43999-5_24}
}

@article{hu2023cross,
  title={CROSS: Cross-domain residual-optimization-based structure strengthening reconstruction for limited-angle CT},
  author={Hu, Dianlin and Zhang, Yikun and Quan, Guotao and Xiang, Jun and Coatrieux, Gouenou and Luo, Shouhua and Coatrieux, Jean-Louis and Ji, Xu and Han, Hongbin and Chen, Yang},
  journal={IEEE Transactions on Radiation and Plasma Medical Sciences},
  volume={7},
  number={5},
  pages={521--531},
  year={2023},
  publisher={IEEE},
  doi={10.1109/TRPMS.2023.3242662}
}

@inproceedings{ayad2024qn,
  title={Qn-mixer: A quasi-newton mlp-mixer model for sparse-view ct reconstruction},
  author={Ayad, Ishak and Larue, Nicolas and Nguyen, Ma{\"\i} K},
  booktitle={Proceedings of the IEEE/CVF conference on computer vision and pattern recognition},
  pages={25317--25326},
  year={2024},
  doi={10.1109/CVPR52733.2024.02392}
}

@inproceedings{chen2017lowb,
  title={Low-dose CT denoising with convolutional neural network},
  author={Chen, Hu and Zhang, Yi and Zhang, Weihua and Liao, Peixi and Li, Ke and Zhou, Jiliu and Wang, Ge},
  booktitle={2017 IEEE 14th international symposium on biomedical imaging (ISBI 2017)},
  pages={143--146},
  year={2017},
  organization={IEEE},
  doi={10.1109/ISBI.2017.7950488}
}

@article{chen2017lowa,
  title={Low-dose CT with a residual encoder-decoder convolutional neural network},
  author={Chen, Hu and Zhang, Yi and Kalra, Mannudeep K and Lin, Feng and Chen, Yang and Liao, Peixi and Zhou, Jiliu and Wang, Ge},
  journal={IEEE transactions on medical imaging},
  volume={36},
  number={12},
  pages={2524--2535},
  year={2017},
  publisher={IEEE},
  doi={10.1109/TMI.2017.2715284}
}

@inproceedings{zhang2021transct,
  title={TransCT: dual-path transformer for low dose computed tomography},
  author={Zhang, Zhicheng and Yu, Lequan and Liang, Xiaokun and Zhao, Wei and Xing, Lei},
  booktitle={International conference on medical image computing and computer-assisted intervention},
  pages={55--64},
  year={2021},
  organization={Springer},
  doi={10.1007/978-3-030-87231-1_6}
}

@inproceedings{wang2021improving,
  title={Improving generalizability in limited-angle CT reconstruction with sinogram extrapolation},
  author={Wang, Ce and Zhang, Haimiao and Li, Qian and Shang, Kun and Lyu, Yuanyuan and Dong, Bin and Zhou, S Kevin},
  booktitle={International Conference on Medical Image Computing and Computer-Assisted Intervention},
  pages={86--96},
  year={2021},
  organization={Springer},
  doi={10.1007/978-3-030-87231-1_9}
}

@article{shao2025multi,
  title={Multi-stage dual-domain progressive network with synergistic training for sparse-view CT reconstruction},
  author={Shao, Jingyuan and Chen, Huabao and Li, Qiankun and Huang, Xiang and Shu, Jiong and Liu, Lingling and Dou, Shaobin and Wang, Hongzhi},
  journal={Neural Networks},
  pages={108221},
  year={2025},
  publisher={Elsevier},
  doi={10.1016/j.neunet.2025.108221}
}

@inproceedings{xia2022transformer,
  title={A transformer-based iterative reconstruction model for sparse-view ct reconstruction},
  author={Xia, Wenjun and Yang, Ziyuan and Zhou, Qizheng and Lu, Zexin and Wang, Zhongxian and Zhang, Yi},
  booktitle={International Conference on Medical Image Computing and Computer-Assisted Intervention},
  pages={790--800},
  year={2022},
  organization={Springer},
  doi={10.1007/978-3-031-16446-0_75}
}

@article{chen2018learn,
  title={LEARN: Learned experts’ assessment-based reconstruction network for sparse-data CT},
  author={Chen, Hu and Zhang, Yi and Chen, Yunjin and Zhang, Junfeng and Zhang, Weihua and Sun, Huaiqiang and Lv, Yang and Liao, Peixi and Zhou, Jiliu and Wang, Ge},
  journal={IEEE transactions on medical imaging},
  volume={37},
  number={6},
  pages={1333--1347},
  year={2018},
  publisher={IEEE},
  doi={10.1109/TMI.2018.2805692}
}

@article{fan2024mvmsrcn,
  title={{MVMS-RCN}: A Dual-Domain Unfolding {CT} Reconstruction with Multi-sparse-view and Multi-scale Refinement-correction},
  author={Fan, Xiaohong and Chen, Ke and Yi, Huaming and Yang, Yin and Zhang, Jianping},
  journal={IEEE Transactions on Computational Imaging},
  year={2024},
  doi={10.1109/TCI.2024.3489040}
}

@article{armato2011lung,
  title={The lung image database consortium ({LIDC}) and image database resource initiative ({IDRI}): a completed reference database of lung nodules on {CT} scans},
  author={Armato III, Samuel G and McLennan, Geoffrey and Bidaut, Luc and McNitt-Gray, Michael F and Meyer, Charles R and Reeves, Anthony P and Zhao, Binsheng and Aberle, Denise R and Henschke, Claudia I and Hoffman, Eric A and others},
  journal={Medical physics},
  volume={38},
  number={2},
  pages={915--931},
  year={2011},
  publisher={Wiley Online Library},
  doi={10.1118/1.3528204}
}

@article{lu2019deeponet,
  title={Deeponet: Learning nonlinear operators for identifying differential equations based on the universal approximation theorem of operators},
  author={Lu, Lu and Jin, Pengzhan and Karniadakis, George Em},
  journal={arXiv preprint arXiv:1910.03193},
  year={2019},
  doi={10.48550/arXiv.1910.03193}
}

\newpage
\appendix
\clearpage
\setcounter{page}{1}

\begin{center}
{\Large\bfseries Resolution-Agnostic Neural Operators for Multi-Rate Sparse-View CT}\\[0.4em]
{\large\bfseries -- Supplementary Material --}\\[1.2em]
Aujasvit Datta$^{1,3\dagger}$\quad
Jiayun Wang$^{1,2\dagger}$\quad
Asad Aali$^{4}$\quad
Anima Anandkumar$^{1}$\\[0.5em]
$^1$Caltech\quad $^2$Georgia Tech\quad $^3$IIT Kanpur\quad $^4$Stanford University\\[0.3em]
{\footnotesize\texttt{aujasvitd22@iitk.ac.in,\ \{peterw,\,anima\}@caltech.edu,\ asadaali@stanford.edu}}\\[0.3em]
{\footnotesize $^\dagger$Equal contribution}
\end{center}
\vspace{1em}

\noindent In this supplementary material, we provide details omitted in the main text, including:
\begin{itemize}%
\item Sec.~\ref{sec:roteq}: more mathematical details of the proposed rotational equivariant DISCO. (Sec.~\ref{sec:rotational_equivalence} ``Sinogram-Space Operator $\NO_s$'' of the main paper.)
\item Sec.~\ref{sec:implement_details}: additional details on experimental setup, specifically sinogram sampling setting,  and implementation details of baseline methods (Sec.~\ref{sec:details} ``Reconstruction with Different Sampling Rates'' of the main paper.)
\item Sec.~\ref{sec:add_results}: more experimental results and visualizations of the method, including numerical results/tables that cannot fit in the main text due to page limit, as well as additional reconstruction visualizations of multiple methods under different sampling rates. (Sec.~\ref{sec:expr} ``Experimental Results'' of the main paper.)
\item Sec.~\ref{sec:compl}: complexity analysis of the proposed method to existing neural architectures, specifically other neural operator architectures and CNNs. (Sec.~\ref{sec:ablation} ``Analysis and Ablation Study'' of the main paper.)
\end{itemize}

\section{Rotational Equivariant  DISCO}
\label{sec:roteq}
We describe the rotational equivariant  DISCO \cite{ocampo2022scalable,liu2024neural} design details below. 

\noindent\textbf{Rotation–equivariant convolution in $(r,\theta)$.}
Parallel-beam CT induces two key symmetries on the sinogram $\p(\theta,r)$:
(i) a rotation of the image by $\phi$ is a \emph{shift in view angle} of the sinogram,
\begin{equation}
\label{eq:rot-shift}
(\p \!\circ\! R_\phi)(\theta,r) \;=\; \p(\theta-\phi,r),
\end{equation}
and (ii) the $\pi$-periodicity-with-flip identity
\begin{equation}
\label{eq:pi-flip}
\p(\theta+\pi,r)\;=\;\p(\theta,-r).
\end{equation}
Define the group action of in-plane rotations $T_\phi$ on sinogram functions by
\begin{equation}
\label{eq:group-action}
(T_\phi f)(\theta,r) := f(\theta-\phi,r),
\quad 
\end{equation}
with the identification 
\begin{align}
    f(\theta+\pi,r)\equiv f(\theta,-r).
\end{align}

Let $\star$ denote a stationary convolution on $(\theta,r)$ (implemented by DISCO with quadrature), 
\begin{equation}
\label{eq:conv}
(f \star \psi)(\theta,r) \;=\; \int_{\mathbb{R}}\!\!\int_{0}^{\pi} 
f(\theta-\tau,\, r-\rho)\,\psi(\tau,\rho)\, d\tau\, d\rho,
\end{equation}
where the domain obeys Eqn.~\eqref{eq:pi-flip}. Then $\star$ is \emph{SO(2)-equivariant} with respect to the action Eqn.~\eqref{eq:group-action}:
\begin{align}
\label{eq:equiv}
\big((T_\phi f)\star \psi\big)(\theta,r)
&= \int\!\!\int f(\theta-\phi-\tau,\,r-\rho)\,\psi(\tau,\rho)\,d\tau\,d\rho\\
&= (f\star \psi)(\theta-\phi,r)\\
&= \big(T_\phi(f\star \psi)\big)(\theta,r).
\end{align}
In the discrete DISCO implementation, \eqref{eq:equiv} holds up to the usual quadrature error.

\paragraph{Padding to realize the symmetry.}
Eqns.~\eqref{eq:rot-shift}–\eqref{eq:equiv} require boundary conditions consistent with Eqn.~\eqref{eq:pi-flip}. 
We therefore use \emph{flipped circular padding} along $\theta$: when indices cross the $\theta=0/\pi$ boundary, we wrap the data while flipping $r$,
\begin{align}
    \p(\theta\!\pm\!\pi, r)\;\mapsto\; \p(\theta, -r),
\end{align}
and apply standard reflective padding along $r$. This enforces physical continuity at $\theta=0$ and $\theta=\pi$, prevents seam artifacts, and makes the DISCO convolutions in $\NO_s$ intrinsically rotation-robust: rotations of $x$ become $\theta$-translations of $p$ (Eqn.~\eqref{eq:rot-shift}), which are naturally handled by the translational equivariance of Eqn.~\eqref{eq:conv}.

\section{Implementation Details}
\label{sec:implement_details}
\subsection{Sinogram Sampling Settings}
\label{sec: sinogram_settings}
For both datasets, we generate ground-truth sinograms using 720 projection angles. For the C4KC-KiTS dataset, we use a parallel-beam geometry with a 256$\times$256 image resolution and a detector size of 300. Projection angles are uniformly sampled over $[0, \pi)$, which is standard for parallel-beam CT. For the AAPM Low-Dose CT dataset, we use a fan-beam geometry with a 256$\times$256 resolution and 672 detector elements, with a source-to-detector distance of 1075. Projection angles are uniformly sampled over $[0, 2\pi)$ to match the acquisition geometry of the dataset. In both datasets, we train and evaluate reconstruction performance at 9-view, 18-view, 36-view, and 72-view acquisition settings by subsampling the corresponding number of projection angles from the 720-view ground-truth sinograms.

\begin{figure*}[t]
    \centering
    \includegraphics[width=1\linewidth]{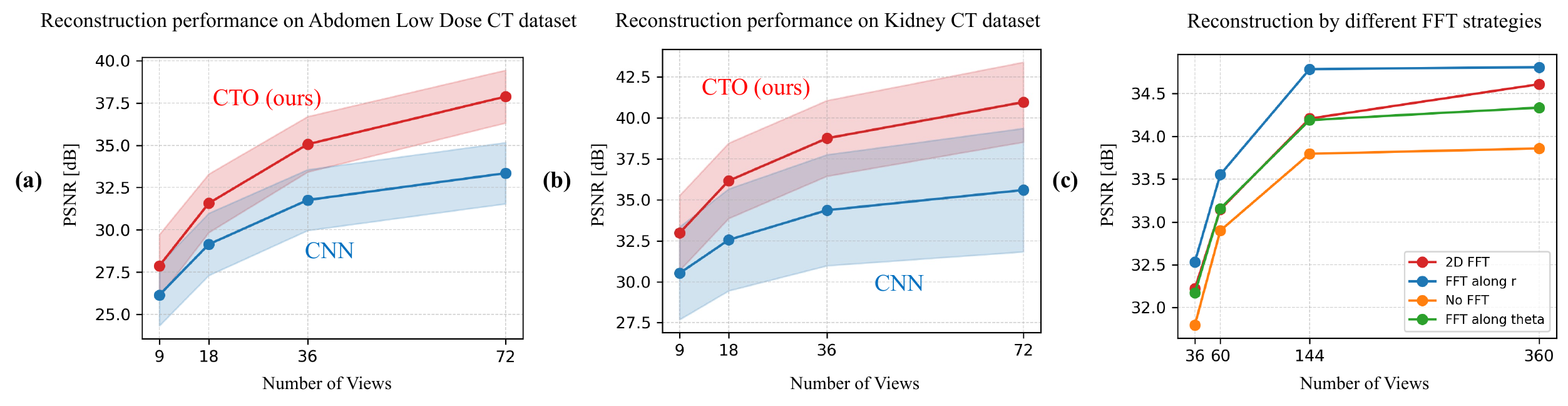}
    \caption{
{\bf (a)} Reconstruction performance on the AAPM abdomen Low-Dose CT dataset \cite{mccollough2016tu} (entire test set). 
{\bf (b)} Reconstruction performance on the Kidney CT dataset \cite{heller2019kits19}  (entire test set). 
In both datasets, \method consistently outperforms the CNN baseline across all sampling rates. 
{\bf (c)} Comparison of reconstruction performance under four frequency-space configurations:
1) 2D FFT: 2D Fourier transform across both axes;
2) FFT along $r$: 1D Fourier transform along the detector (\(r\)) axis;
3) No FFT: no Fourier transform (spatial-domain learning only).
4) FFT along $\theta$: 1D Fourier transform along the angular (\(\theta\)) axis;
Introducing frequency-space processing yields consistent improvements over the spatial-only baseline, with detector-axis transforms providing the largest gains, in line with the analysis in Sec.~\ref{sec:rotational_equivalence}. \\
}
    \label{fig:nos_design}
\end{figure*}

\subsection{Baseline Setup}
For fair comparisons, \method and baseline methods all use the same  amount and configurations of training data.
\begin{enumerate}
    \item \textbf{Learning-free iterative methods.} We include two classical reconstruction baselines: Filtered Back-Projection (FBP) \cite{shepp1974fourier} and the Simultaneous Algebraic Reconstruction Technique (SART) \cite{andersen1984simultaneous}. FBP reconstruction applies the inverse Radon transform configured as in Sec.~\ref{sec: sinogram_settings} on the subsampled sinograms after passing them through a ramp frequency filter. For SART, we use the public \texttt{scikit-image} \texttt{iradon\_sart} operator. For each test image, we perform one initial SART sweep followed by four additional refinement iterations (five total). The relaxation parameter is left at the default internal setting of \texttt{iradon\_sart}. Following standard conventions, all SART updates are clipped to the valid attenuation range [0, 0.549] to avoid divergence.
    \item \textbf{DuDoTrans}. DuDoTrans \cite{wang2022dudotrans} is a dual-domain sparse-view CT reconstruction framework that jointly enhances sinograms and images. A Sinogram Restoration Transformer (SRT) models long-range dependencies in the projection domain to restore informative sinograms, which are converted to an intermediate image through a differentiable consistency layer and fused with an initial FBP reconstruction. A Residual Image Reconstruction Module then refines the fused representation to produce the final output. We use the official implementation, training for 100 epochs with Adam (initial learning rate $1\mathrm{e}{-4}$) and a step decay (learning rate multiplied by $0.1$ after epoch 20).
    \item \textbf{GloReDi}. GloReDi \cite{li2023learning} is an image-domain framework for sparse-view CT that distills global representations (GloRe) from intermediate-view reconstructions to improve sparse-view quality. The teacher network (intermediate) guides the student (sparse) through EMA-updated weights. We use the official implementation, training for 100 epochs with Adam ($\alpha=1e-4$).
    \item \textbf{MDPRNet}. MDPRNet \cite{shao2025multi} is a multi-stage dual-domain progressive reconstruction network for sparse-view CT. A U-shaped subnetwork first restores the sparse sinogram, which is converted to an intermediate image via a differentiable FBP layer. A second U-shaped subnetwork refines this image, and a final single-scale feature subnetwork (SSFNet) operates at the original resolution to preserve fine spatial details. Cross-stage Feature Adapters (CFA), comprising a Learnable Global Attention Gate and a Supervised Attention Module, regulate information flow between stages. The architecture further employs a Multi-view Synergistic Training Strategy (MSTS) that groups sparse-view settings into ultra-sparse and sparse-view categories, enabling a single model to handle multiple view counts. We use the official implementation of the method.
    \item \textbf{Learned Primal-Dual (LPD)}. LPD \cite{adler2018learned} is a learned iterative reconstruction scheme that unrolls the primal-dual hybrid gradient (PDHG) algorithm, replacing the proximal operators with convolutional neural networks. Each unrolled iteration maintains extended primal and dual variables ($N_{\text{primal}}{=}N_{\text{dual}}{=}5$ channels) and applies a three-layer residual CNN (with $3\times3$ convolutions, 32 hidden channels, and PReLU activations) as the learned proximal in both primal and dual spaces. The forward operator and its adjoint are evaluated once per iteration, connecting the two domains. The algorithm works directly from raw projection data with zero-initialization, requiring no initial FBP reconstruction. We use the official implementation with $I{=}10$ unrolled iterations (60 total convolutional layers, ${\sim}2.4 \times 10^5$ parameters), trained for $10^5$ batches using Adam (initial learning rate $1\mathrm{e}{-3}$ with cosine annealing, $\beta_2{=}0.99$) and an MSE loss with gradient norm clipping set to 1.
    \item \textbf{LEARN}. LEARN \cite{chen2018learn} is an unrolled iterative reconstruction network that casts the fields-of-experts (FoE) regularized optimization into a deep feed-forward architecture. Each iteration block comprises a data-fidelity gradient step $\lambda^t A^\top(Ax^t - y)$ and a three-layer CNN acting as the learned regularizer, connected via a residual shortcut. All regularization terms and balancing parameters are iteration-dependent and learned end-to-end from projection–image pairs. The network uses 48 filters of size $5\times5$ in the first two convolutional layers and a single output filter per iteration, with the FBP result as initialization. We use the official implementation with $N_t{=}30$ unrolled iterations (90 total layers), trained with the Adam optimizer (initial learning rate $1\mathrm{e}{-4}$, decayed to $1\mathrm{e}{-5}$) using an accumulated MSE loss.
    \item \textbf{RegFormer}. RegFormer \cite{xia2022transformer} is a transformer-based unrolled iterative reconstruction network that learns both local and nonlocal regularization for sparse-view CT. The gradient descent iteration is unrolled into alternating local and nonlocal iteration blocks: local blocks use three CNN layers with $5\times5$ kernels for artifact suppression, while nonlocal blocks employ two successive Swin Transformer blocks to capture long-range dependencies for structure preservation. An FBP operator replaces the standard back-projection in the data-fidelity step to accelerate convergence within the fixed iteration budget. An iteration transmission (IT) module passes intermediate feature maps between consecutive blocks via skip connections, enabling deeper feature extraction across iterations. We use the official implementation with $N_t{=}18$ iteration blocks and $C{=}96$ feature channels, trained for 200 epochs with AdamW (initial learning rate $1\mathrm{e}{-4}$, decayed to 0).
    \item \textbf{Unrolled networks with CNNs.} We use an unrolled U-Net \cite{ronneberger2015u} baseline composed of 3 image space cascades, where each cascade consists of a U-Net refinement module followed by a data consistency update. This baseline operates entirely in the image space and does not apply any learned processing in the sinogram domain. The U-Net uses a standard encoder–decoder configuration with 32 hidden channels and four pooling levels. The model is trained under the same loss function and optimization settings as our method to ensure a fair comparison.
    \item \textbf{Diffusion models}. Diffusion probabilistic models (DPMs) have been shown as robust methods for solving ill-posed inverse problems. In such inverse problems, we aim to recover $x_0$ from corrupted measurements $y$, requiring sampling from the posterior $p(x_0 | y)$. This stochastic process is modeled by a stochastic differential equation (SDE) \cite{song2020score, karras2022elucidating}: $dx = -2\dot{\sigma}(t)(\mathbb{E}[x_0 | x_t, y] - x_t) dt + g(t) dw.$ Because computing likelihood $p(y | x_t)$ is computationally intractable, we use approximations, such as Diffusion Posterior Sampling (DPS) \cite{chung2023dps} and Annealed Langevin Dynamics (ALD) \cite{jalal2021robust}, that leverage DPMs as priors for accelerated reconstruction and approximate $p(y | x_t)$ as $p(y | x_0=\mathbb{E}[x_0 | x_t])$. For the diffusion baseline, we train a DPM following the EDM \cite{karras2022elucidating} loss formulation. The model is trained for $2,000$ iterations, employing a UNet-style architecture comprising $65$ million trainable parameters, where the learning rate, noise schedule, and training were set following \cite{karras2022elucidating}. For inference, we utilize a $1,000$-step noise schedule (data-consistency weighting parameter $\gamma=5e-3$), and use the trained DPM to perform reconstruction across both approximation methods: (i) DPS \cite{chung2023dps} and (ii) ALD \cite{jalal2021robust}. 
\end{enumerate}

\subsection{Hardware and Training.}
The training of our model and baselines is with a batch size of 8 across 4 A100 (40GB) GPUs.

\section{Additional Results and Visualizations}
\label{sec:add_results}

\subsection{Kidney CT Dataset Results}
\label{sec:kidney}

\begin{table*}[t]
\centering
\setlength{\tabcolsep}{5pt}
\renewcommand{\arraystretch}{1.2}
\caption{Sparse-view CT reconstruction results (including diffusion models) for smaller test set of 500 images from the kidney CT dataset  \cite{heller2019kits19}. Results on entire test set of 10,000 images are provided in Table~\ref{tab:kits_multires_results_full}. Lower is better for RMSE; higher is better for PSNR/SSIM.}
\vspace{-0.5em}
\label{tab:kits_multires_results}
\resizebox{\linewidth}{!}{%
\begin{tabular}{llccccccccc}
\toprule
\multirow{2}{*}{Category} & \multirow{2}{*}{Method} &
\multicolumn{3}{c}{18-view} & \multicolumn{3}{c}{36-view} & \multicolumn{3}{c}{72-view} \\
\cmidrule(lr){3-5}\cmidrule(lr){6-8}\cmidrule(lr){9-11}
& & RMSE (HU)$\downarrow$ & PSNR$\uparrow$ & SSIM (x$10^{-2}$) $\uparrow$
  & RMSE (HU)$\downarrow$ & PSNR$\uparrow$ & SSIM (x$10^{-2}$) $\uparrow$
  & RMSE (HU)$\downarrow$ & PSNR$\uparrow$ & SSIM (x$10^{-2}$) $\uparrow$ \\
\midrule
\multirow{2}{*}{\begin{tabular}{c}Learning-free\end{tabular}}
& FBP \cite{shepp1974fourier}     & 1541.70 $\pm$ 190.59 & 6.74 $\pm$ 1.45 & 5.00 $\pm$ 3.00 & 1502.42 $\pm$ 185.69 & 6.96 $\pm$ 1.45 & 8.78 $\pm$ 2.81 & 1423.67 $\pm$ 175.85 & 7.43 $\pm$ 1.45 & 16.54 $\pm$ 2.48 \\
& SART \cite{andersen1984simultaneous}  & 678.26 $\pm$ 140.11  & 14.13 $\pm$ 3.09 & 48.93 $\pm$ 6.60 & 678.86 $\pm$ 144.47 & 14.16 $\pm$ 3.27 & 50.93 $\pm$ 7.61 & 675.58 $\pm$ 145.43  & 14.22 $\pm$ 3.35 & 51.00 $\pm$ 8.84 \\
\midrule
\multirow{2}{*}{\begin{tabular}{c}Diffusion\end{tabular}}
& DPS  \cite{chung2023dps}           & 81.67 $\pm$ 21.72 & 32.50 $\pm$ 2.45 & 88.03 $\pm$ 4.10 & 47.25 $\pm$ 8.73 & 37.08 $\pm$ 1.80 & 94.38 $\pm$ 1.63 & 41.95 $\pm$ 6.99 & 38.08 $\pm$ 1.70 & 95.34 $\pm$ 1.53 \\

& ALD    \cite{jalal2021robust}       & 92.26 $\pm$ 16.09  & 31.25 $\pm$ 1.72 & 85.58 $\pm$ 3.15 & 53.21 $\pm$ 8.97 & 36.02 $\pm$ 1.63 & 93.27 $\pm$ 1.80 & 43.28 $\pm$ 6.70  & 37.79 $\pm$ 1.59 & 95.27 $\pm$ 1.53 \\
\midrule
\multirow{3}{*}{\begin{tabular}{c}Single pass\end{tabular}}
& DuDoTrans  \cite{wang2022dudotrans}  & 111.86 $\pm$ 25.87  & 29.62 $\pm$ 2.07 & 84.64 $\pm$ 5.19 & 64.69 $\pm$ 16.01 & 34.42 $\pm$ 2.18 & 90.94 $\pm$ 3.85 & 48.03 $\pm$ 13.06 & 37.04 $\pm$ 2.37 & 94.49 $\pm$ 3.19 \\
& GloReDi    \cite{li2023learning}     & 108.36 $\pm$ 84.92  & 30.76 $\pm$ 3.86 & 85.28 $\pm$ 7.42 & 94.30 $\pm$ 86.62 & 32.19 $\pm$ 4.09 & 87.76 $\pm$ 7.11 & 84.60 $\pm$ 88.21  & 33.37 $\pm$ 4.32 & 90.01 $\pm$ 7.15 \\
& MDPRNet    \cite{shao2025multi}     & 119.52 $\pm$ 27.41  & 27.57 $\pm$ 2.43 & 84.97 $\pm$ 6.44 & 82.76 $\pm$ 20.17 & 33.42 $\pm$ 1.96 & 87.47 $\pm$ 6.41 & 65.43 $\pm$ 6.79  & 34.67 $\pm$ 2.64 & 89.27 $\pm$ 4.46 \\
\midrule
\multirow{5}{*}{\begin{tabular}{c}Unrolled Networks\end{tabular}}
& Learned PD \cite{adler2018learned} & 140.94 $\pm$ 20.47  & 27.52 $\pm$ 1.58 & 71.66 $\pm$ 6.40 & 106.43 $\pm$ 15.78 & 29.96 $\pm$ 1.60 & 77.51 $\pm$ 4.94 & 76.17 $\pm$ 13.49  & 32.90 $\pm$ 1.77 & 86.49 $\pm$ 3.45 \\
& LEARN  \cite{chen2018learn} & 122.01 $\pm$ 25.14  & 28.86 $\pm$ 1.83 & 83.25 $\pm$ 4.22 & 86.27 $\pm$ 17.76 & 31.87 $\pm$ 1.85 & 88.33 $\pm$ 3.25 & 62.22 $\pm$ 14.49  & 34.75 $\pm$ 2.02 & 92.52 $\pm$ 2.45 \\
& RegFormer\cite{xia2022transformer}  & 119.03 $\pm$ 84.37  & 29.86 $\pm$ 3.82 & 84.47 $\pm$ 7.24 & 88.62 $\pm$ 87.77 & 32.94 $\pm$ 4.38 & 89.08 $\pm$ 6.46 & 69.71 $\pm$ 59.65  & 35.68 $\pm$ 4.90 & 92.56 $\pm$ 6.01 \\
& Unrolled CNN  \cite{modl_19} & 82.52 $\pm$ 41.09  & 32.66 $\pm$ 2.82 & 89.54 $\pm$ 3.33 & 67.67 $\pm$ 38.30 & 34.53 $\pm$ 3.11 & 91.92 $\pm$ 3.12 & 60.65 $\pm$ 43.93  & 35.82 $\pm$ 3.64 & 93.84 $\pm$ 2.95 \\
& \textbf{CTO (ours)} & \textbf{53.50 $\pm$ 13.55} & \textbf{36.11 $\pm$ 2.25} & \textbf{92.56 $\pm$ 2.50} & \textbf{39.67 $\pm$ 9.59} & \textbf{38.68 $\pm$ 2.20} & \textbf{94.85 $\pm$ 1.85} & \textbf{31.07 $\pm$ 8.11} & \textbf{40.83 $\pm$ 2.36} & \textbf{96.42 $\pm$ 1.55} \\
\bottomrule
\end{tabular}
;}
\vspace{-1em}
\end{table*}

\begin{table*}[t]
\centering
\setlength{\tabcolsep}{5pt}
\renewcommand{\arraystretch}{1.2}
\caption{Sparse-view CT reconstruction results for Kidney CT dataset \cite{heller2019kits19} for the larger test set of 10,000 images. Diffusion models are omitted due to high computational costs. Lower is better for RMSE; higher is better for PSNR/SSIM.}
\vspace{-0.5em}
\label{tab:kits_multires_results_full}
\resizebox{\linewidth}{!}{%
\begin{tabular}{llccccccccc}
\toprule
\multirow{2}{*}{Category} & \multirow{2}{*}{Method} &
\multicolumn{3}{c}{18-view} & \multicolumn{3}{c}{36-view} & \multicolumn{3}{c}{72-view} \\
\cmidrule(lr){3-5}\cmidrule(lr){6-8}\cmidrule(lr){9-11}
& & RMSE (HU)$\downarrow$ & PSNR$\uparrow$ & SSIM (x$10^{-2}$) $\uparrow$
  & RMSE (HU)$\downarrow$ & PSNR$\uparrow$ & SSIM (x$10^{-2}$) $\uparrow$
  & RMSE (HU)$\downarrow$ & PSNR$\uparrow$ & SSIM (x$10^{-2}$) $\uparrow$ \\
\midrule
\multirow{2}{*}{\begin{tabular}{c}Learning-free\end{tabular}}
& FBP \cite{shepp1974fourier}     & 1539.09 $\pm$ 198.54 & 6.83 $\pm$ 1.89 & 5.20 $\pm$ 3.46 & 1499.88 $\pm$ 193.44 & 7.05 $\pm$ 1.89 & 8.97 $\pm$ 3.27 & 1421.26 $\pm$ 183.22 & 7.52 $\pm$ 1.89 & 16.72 $\pm$ 2.94 \\
& SART \cite{andersen1984simultaneous}  & 675.36 $\pm$ 142.01  & 14.24 $\pm$ 3.30 & 49.08 $\pm$ 7.84 &  675.91 $\pm$ 146.27 & 14.27 $\pm$ 3.47 & 51.08 $\pm$ 7.94 &  672.58 $\pm$ 147.23  & 14.33 $\pm$ 3.54 & 51.19 $\pm$ 9.16 \\
\midrule
\multirow{3}{*}{\begin{tabular}{c}Single pass\end{tabular}}
& DuDoTrans  \cite{wang2022dudotrans}  & 112.77 $\pm$ 27.57  & 29.63 $\pm$ 2.14 & 84.61 $\pm$ 5.32 & 65.34 $\pm$ 17.20 & 34.42 $\pm$ 2.25 & 90.88 $\pm$ 3.99 & 48.52 $\pm$ 14.22 & 37.05 $\pm$ 2.40 & 94.41 $\pm$ 3.26 \\
& GloReDi    \cite{li2023learning}     & 111.95 $\pm$ 92.06  & 30.68 $\pm$ 4.04 & 85.08 $\pm$ 7.81 & 97.92 $\pm$ 93.81 & 32.10 $\pm$ 4.28 & 87.57 $\pm$ 7.51 & 88.50 $\pm$ 95.46  & 33.24 $\pm$ 4.52 & 89.74 $\pm$ 7.54 \\
& MDPRNet    \cite{shao2025multi}     & 117.46 $\pm$ 25.57  & 27.80 $\pm$ 2.49 & 84.98 $\pm$ 8.28 & 81.43 $\pm$ 19.73 & 33.37 $\pm$ 2.04 & 88.16 $\pm$ 6.14 & 66.13 $\pm$ 6.29  & 35.04 $\pm$ 2.78 & 89.42 $\pm$ 3.97 \\
\midrule
\multirow{3}{*}{\begin{tabular}{c}Unrolled Networks\end{tabular}}
& Learned PD  \cite{adler2018learned} & 141.94 $\pm$ 21.43  & 27.53 $\pm$ 1.69 & 71.66 $\pm$ 6.60 & 107.34 $\pm$ 16.87 & 29.97 $\pm$ 1.68 & 77.47 $\pm$ 5.17 & 76.71 $\pm$ 14.24  & 32.92 $\pm$ 1.83 & 86.45 $\pm$ 3.59 \\
& LEARN\cite{chen2018learn} & 123.22 $\pm$ 26.74  & 28.85 $\pm$ 1.91 & 83.22 $\pm$ 4.35 & 87.28 $\pm$ 20.80 & 31.87 $\pm$ 1.93 & 88.30 $\pm$ 3.36 & 63.15 $\pm$ 18.87  & 34.75 $\pm$ 2.04 & 92.48 $\pm$ 2.50 \\
& RegFormer\cite{xia2022transformer}  & 123.05 $\pm$ 91.37  & 29.76 $\pm$ 3.97 & 84.29 $\pm$ 7.45 & 92.43 $\pm$ 85.41 & 32.84 $\pm$ 4.57 & 88.88 $\pm$ 6.77 & 73.55 $\pm$ 68.62  & 35.55 $\pm$ 5.14 & 92.34 $\pm$ 6.35 \\
& Unrolled CNN  \cite{modl_19} & 82.51 $\pm$ 48.28  & 32.56 $\pm$ 3.11 & 89.48 $\pm$ 3.50 & 70.74 $\pm$ 45.33 & 34.37 $\pm$ 3.39 & 91.82 $\pm$ 3.35 & 63.13 $\pm$ 45.85  & 35.60 $\pm$ 3.77 & 93.71 $\pm$ 3.26 \\
& \textbf{CTO (ours)} & \textbf{53.60 $\pm$ 13.81} & \textbf{36.17 $\pm$ 2.30} & \textbf{92.63 $\pm$ 2.55} & \textbf{39.71 $\pm$ 10.05} & \textbf{38.76 $\pm$ 2.31} & \textbf{94.93 $\pm$ 1.88} & \textbf{30.85 $\pm$ 8.08} & \textbf{40.97 $\pm$ 2.44} & \textbf{96.49 $\pm$ 1.50} \\
\bottomrule
\end{tabular}
}
\end{table*}

\noindent \textbf{Subset results.} Due to the long inference time and computation of diffusion methods, we report all methods (including diffusion ones) on a 500-slice subset. 

\noindent \textbf{Full-set results.} We also provide results of all methods excluding diffusion ones on the entire test set of 10,000 images for the Kidney CT dataset \cite{heller2019kits19} in Table~\ref{tab:kits_multires_results_full}. We plot the numerical results in Fig.~\ref{fig:nos_design}b. 

\noindent We also plot results on AAPM dataset in Fig.~\ref{fig:nos_design}a.

\subsection{Rotational Equivariance Results}

We perform an ablation study to evaluate the impact of circular padding on reconstruction quality. As discussed in Sec.~\ref{sec:rotational_equivalence}, circular padding enforces the periodic boundary conditions inherent to the sinogram’s angular domain and enables the network to better exploit rotational equivariance. Table~\ref{tab:circular_padding} shows the reconstruction performance with and without circular padding under a 24-view  subsampling setting. Incorporating circular padding leads to a notable improvement in PSNR, indicating that handling angular periodicity explicitly enhances the overall reconstruction quality.

\begin{table}[h]
\centering
\caption{Ablation study demonstrating the effect of the proposed rotational equivariant DISCO (via circular padding) on reconstruction quality.}
\label{tab:circular_padding}
\resizebox{0.7\linewidth}{!}{\begin{tabular}{l c}
\toprule
Method & PSNR on 24-view subsampled reconstruction (dB) \\
\midrule
CTO  & \textbf{33.00} \\
CTO (no Rot. Eqv.) & 32.17 \\
\bottomrule
\end{tabular}
}
\vspace{2mm}
\end{table}

\subsection{Sinogram Space $\NO_s$ Design Variants}
\label{sec:supp_nos}
To determine the most effective configuration for frequency-space learning, we evaluate different Fourier transform strategies for the sinogram prior to operator learning by $\NO_{s, \text{freq}}$. Specifically, we consider four variants:
   1) \textbf{Detector-axis transform:} one-dimensional Fourier transform applied along the detector ($r$) axis.
    2) \textbf{Angle-axis transform:} one-dimensional Fourier transform applied along the angular ($\theta$) axis.
    3) \textbf{2D transform:} two-dimensional Fourier transform applied along both detector and angular axes.
    4) \textbf{Baseline (no transform):} no frequency-space learning is applied, and the network operates only in the spatial domain of the sinogram and image.

\noindent The comparative results of these configurations are summarized in Supplementary-Fig.~\ref{fig:nos_design}. We observe that incorporating frequency-space learning in any form leads to a consistent boost in reconstruction quality compared to purely spatial-domain learning (at least 1 dB PSNR gain). Among the variants, applying the Fourier transform along the detector axis provides the largest performance gain (2.7 dB PSNR gain), consistent with the theoretical advantages of this strategy as detailed in Sec.~\ref{sec:rotational_equivalence}.

\begin{figure*}[t]
    \centering
    \includegraphics[width=\linewidth]{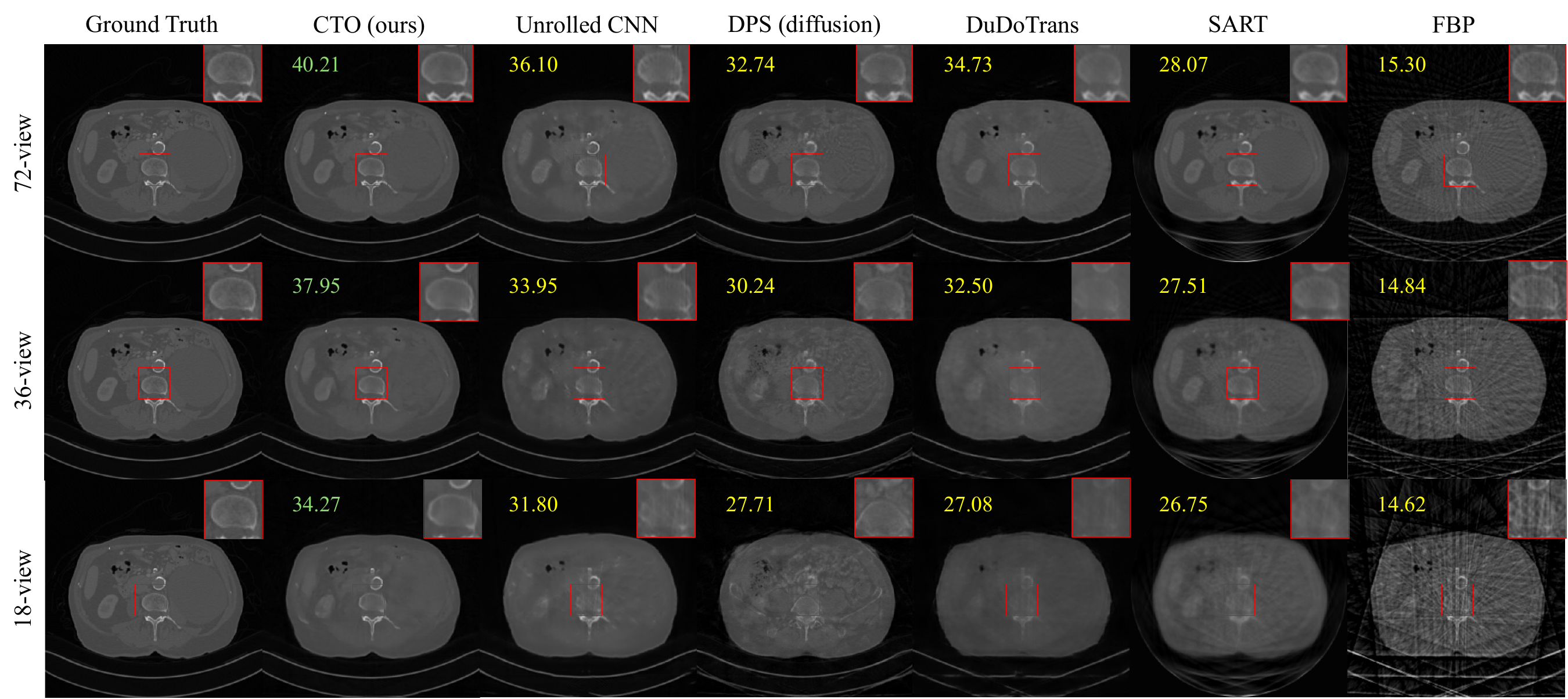}
    \caption{Reconstruction on multiple sampling rates for Abdomen Low-Dose CT dataset \cite{mccollough2016tu}. \method outperforms baseline methods for PSNR (upper left) and reconstruction quality across all sinogram sampling rates.}
    \label{fig:multi_acc_rate_vis}
\end{figure*}

\subsection{DISCO Hyperparameter Analysis}
\label{sec:DISCOhyper}

Hyperparameter optimization was performed using Bayesian search over six 
configuration parameters: basis type (piecewise linear, Morlet, Zernike), 
number of basis functions (\texttt{num\_basis\_func}), number of bases 
(\texttt{num\_basis}), and radial cutoff (\texttt{radius\_cutoff}). Parameter 
importance was estimated using a random forest regressor trained on the swept 
configurations with validation PSNR as the target metric, with feature 
importances serving as a proxy for predictive relevance. Linear association 
between each parameter and validation PSNR was additionally quantified via 
Pearson correlation.

As summarized in Table~\ref{tab:hyperparam}, \texttt{basis\_type.piecewise\_linear} 
emerged as the most important predictor of reconstruction quality, followed by 
\texttt{num\_basis\_func} and \texttt{num\_basis}. While Morlet and Zernike basis 
types showed negligible importance scores, both exhibited moderate negative 
correlation with validation PSNR, indicating that their selection is actively 
detrimental rather than merely uninformative. These findings guided our final 
model configuration, with piecewise linear basis functions selected as the default.

\begin{table}[t]
\centering
\caption{Hyperparameter importance and correlation with validation PSNR. Importance scores were computed by training a random forest regressor with the swept hyperparameters as input features and validation PSNR as the target, reporting each feature's importance as a measure of its predictive relevance. Correlation values report the Pearson linear correlation between each hyperparameter and validation PSNR, ranging from -1 to 1. Importance captures nonlinear and interaction effects between hyperparameters, whereas correlation isolates linear associations of individual parameters.}
\label{tab:hyperparam}
\begin{tabular*}{0.65\textwidth}{l @{\extracolsep{\fill}} cc}
\toprule
\textbf{Config Parameter} & \textbf{Importance} & \textbf{Correlation} \\
\midrule
\texttt{basis\_type}: piecewise linear & 0.429 & 0.468   \\
\texttt{num\_basis\_func}              & 0.278 & 0.032 \\
\texttt{num\_basis}                    & 0.145    & -0.176     \\
\texttt{radius\_cutoff}               & 0.139   & 0.214 \\
\texttt{basis\_type}: Morlet          & 0.008 & -0.412 \\
\texttt{basis\_type}: Zernike         & 0.002 & -0.196 \\
\bottomrule
\end{tabular*}
\end{table}

\subsection{Out-of-Distribution Generalization}
\label{sec:suppOOD}

\noindent \textbf{Generalization on noisy sinogram data.} We evaluate our zero-noise trained CTO and unrolled CNN on sinogram data with added Poisson noise ($I_0=10^6$). On average over different subsampling rates (9, 18, 36, 72-view), CTO outperforms unrolled CNN by 3.06 dB PSNR. This shows that CTO generalizes better to noisy conditions observed in real clinical settings.

\noindent \textbf{Generalization across different datasets.} To evaluate the ability of our model to generalize across different datasets, we test the CTO and unrolled CNN models trained on C4KC-KiTS dataset on the AAPM dataset. We observe that CTO performs substantially better than unrolled CNN with an average gain of 9.8 dB PSNR across different subsampling rates.

\subsection{9-View AAPM Test Results}
\label{sec:9view}
We provide test results for 9-view SVCT reconstruction on the AAPM dataset \cite{mccollough2016tu} in Table~\ref{tab:aapm_9x_results}.
\begin{table}[htbp]
\centering
\setlength{\tabcolsep}{5pt}
\renewcommand{\arraystretch}{1.2}
\caption{9-view CT reconstruction results for AAPM
dataset \cite{mccollough2016tu}. Lower is better for RMSE; higher is better for PSNR/SSIM. For all learning methods, a single model is trained on multi-view sinograms with $N_v = {9, 18, 36, 72}$ views together. For fairness, all methods are trained using the same amount of data -- this differs from the resolution-specific models trained in their original paper since multi-resolution training generally reduces overfitting (Table~\ref{tab:combined}a). Test results for the rest of the rates are reported in main text Table~\ref{tab:aapm_multires_results}.}
\label{tab:aapm_9x_results}
\resizebox{0.8\textwidth}{!}{%
\begin{tabular}{llccc}
\toprule
\multirow{2}{*}{{\bf Category}} & \multirow{2}{*}{{\bf Method}} &
\multicolumn{3}{c}{{\bf 9-view}} \\
\cmidrule(lr){3-5}
& & RMSE (HU)$\downarrow$ & PSNR$\uparrow$ & SSIM (x$10^{-2}$) $\uparrow$ \\
\midrule
\multirow{2}{*}{\begin{tabular}{c}Learning-free\end{tabular}}
& FBP \cite{shepp1974fourier}    & 640.64 $\pm$ 42.73 & 13.03 $\pm$ 1.63  & 33.29 $\pm$ 6.09 \\
& SART \cite{andersen1984simultaneous}  & 681.87 $\pm$ 130.27 & 14.07 $\pm$ 2.88 & 45.31 $\pm$ 6.46 \\
\midrule
\multirow{3}{*}{\begin{tabular}{c}Single pass\end{tabular}}
& DuDoTrans \cite{wang2022dudotrans} & 350.61 $\pm$ 18.22 & 18.26 $\pm$ 1.60 & 48.64 $\pm$ 4.91 \\
& GloReDi \cite{li2023learning}        & 159.11 $\pm$ 86.01 & 27.12 $\pm$ 3.45 & 78.30 $\pm$ 8.07 \\
& MDPRNet \cite{shao2025multi}        & 346.89 $\pm$ 22.17 & 20.81 $\pm$ 1.67 & 64.57 $\pm$ 6.40 \\
\midrule
\multirow{5}{*}{\begin{tabular}{c}Unrolled Networks\end{tabular}}
& Learned PD \cite{adler2018learned} & 203.95 $\pm$ 21.92 & 22.99 $\pm$ 1.73 & 39.87 $\pm$ 5.11 \\
& LEARN \cite{chen2018learn} & 205.77 $\pm$ 25.54 & 22.94 $\pm$ 1.73 & 61.49 $\pm$ 4.29 \\
& RegFormer \cite{xia2022transformer} & 196.114 $\pm$ 23.68 & 23.35 $\pm$ 1.79 & 65.52 $\pm$ 4.84 \\
& Unrolled CNN \cite{modl_19} & 142.45 $\pm$ 20.73 & 26.14 $\pm$ 1.82 & 68.99 $\pm$ 8.37 \\
& \textbf{CTO (ours)} & \textbf{116.69 $\pm$ 15.86} & \textbf{27.87 $\pm$ 1.85} & \textbf{74.35 $\pm$ 4.98} \\
\bottomrule
\end{tabular}
}
\end{table}
\subsection{Multi-Rate Visualization}
Fig.~\ref{fig:multi_acc_rate_vis} shows reconstructions on the AAPM abdomen low-dose CT dataset \cite{mccollough2016tu} across multiple sampling rates. CTO preserves reconstruction quality even under heavy undersampling with noticeably fewer artifacts compared to other methods.

\subsection{Single Rate training and inference}

To further analyze the performance of our model, we also train/test per rate as an ablation. This differs from our default setting, where a single model is trained jointly across all sampling rates and must generalize across them at test time; here, instead, a separate model is trained and evaluated separately on each fixed rate (18, 36, or 72 views), which is the default setting most other methods are designed and evaluated on. Under this setting, CTO still outperforms all baselines at every view count (Table~\ref{tab:multirate_fairness}), and the margin widens further under our default multi-rate regime.

\begin{table*}[b]
\centering
\renewcommand{\arraystretch}{1.2}
\caption{Single-rate vs.\ multi-rate training comparison (PSNR$\uparrow$, dB). Multi-rate is the setting we study: a single model is trained jointly across all sampling rates and must generalize across them at test time, whereas single-rate trains and tests a separate model per rate. CTO outperforms all baselines under both protocols, and the margin of improvement widens in the Multi-rate training setting. The results for multi-rate training are same as those in Table~\ref{tab:aapm_multires_results}}
\label{tab:multirate_fairness}
\resizebox{0.8\textwidth}{!}{
\begin{tabular*}{\textwidth}{l @{\extracolsep{\fill}} ccc | ccc}
\toprule
& \multicolumn{3}{c}{\textbf{Single rate training}} & \multicolumn{3}{c}{\textbf{Multi rate training}} \\
\cmidrule(lr){2-4} \cmidrule(lr){5-7}
\textbf{Method (PSNR$\uparrow$)} & \textbf{18-view} & \textbf{36-view} & \textbf{72-view} & \textbf{18-view} & \textbf{36-view} & \textbf{72-view} \\
\midrule
LEARN        & 27.97 & 30.76 & 32.32 & 25.51 & 27.53 & 29.90 \\
RegFormer    & 29.47 & 32.18 & 35.11 & 25.67 & 28.00 & 30.43 \\
Unrolled CNN & 32.35 & 35.63 & 38.35 & 29.14 & 31.76 & 33.35 \\
\textbf{CTO (ours)} & \textbf{33.94} & \textbf{36.26} & \textbf{38.83} & \textbf{31.57} & \textbf{35.06} & \textbf{37.88} \\
\bottomrule
\end{tabular*}
}
\end{table*}

\subsection{LIDC-IDRI Lung CT Dataset Results}
\label{sec:lidc}

We also compare \method and Unrolled CNN on the LIDC-IDRI dataset \cite{armato2011lung} under the same multi-rate training protocol as the main paper (a single model trained jointly across 18-, 36-, and 72-view rates). LIDC-IDRI is a publicly available lung CT benchmark of 1,018 thoracic scans from 1,010 patients. Results are in Table~\ref{tab:lidc_results}. \method outperforms Unrolled CNN across all rates, with gains of $+1.54$\,dB, $+2.47$\,dB, and $+3.29$\,dB PSNR at 18-, 36-, and 72-view respectively.

\begin{table}[h]
\centering
\caption{Sparse-view CT reconstruction on the LIDC-IDRI lung CT dataset \cite{armato2011lung}. Higher is better for PSNR\,/\,SSIM.}
\label{tab:lidc_results}
\resizebox{\linewidth}{!}{\begin{tabular}{lccc}
\toprule
\textbf{Method} & \textbf{18-view (PSNR\,/\,SSIM)} & \textbf{36-view (PSNR\,/\,SSIM)} & \textbf{72-view (PSNR\,/\,SSIM)} \\
\midrule
Unrolled CNN & 33.89 / 0.889 & 35.11 / 0.914 & 36.38 / 0.941 \\
\textbf{\method} & \textbf{35.43} / \textbf{0.901} & \textbf{37.58} / \textbf{0.932} & \textbf{39.67} / \textbf{0.960} \\
\bottomrule
\end{tabular}}
\end{table}

\section{Complexity Analysis}
\label{sec:compl}
\subsection{Computational Complexity of Neural Operators}
\label{sec:comp_NOs}

We analyze the per-layer computational cost of four neural operators on an $N \times N$ grid with $c$ hidden channels. FNO~\cite{fnop} retains $k_{\max}$ Fourier modes per spatial axis for its spectral convolution. DISCO~\cite{liu2024neural} uses a compactly supported kernel with stencil size $s$ (the number of spatial neighbors within the support radius). DeepONet~\cite{lu2019deeponet} parameterizes its branch and trunk MLPs with hidden width $d$, basis rank $p$, and depth $L$. LocalNO~\cite{liu2024neural} augments FNO with two parallel local branches---a differential kernel and a DISCO convolution---summed pointwise per layer. For a fair comparison, we set $c = 64$, $N = 256$, $k_{\max} = 256$ (no spectral truncation, giving FNO maximum expressivity), $s = 9$ (a $3 \times 3$ local stencil), $d = 128$, $p = 64$, and $L = 4$.

\begin{table}[b]
\centering
\caption{Per-layer computational complexity of neural operators at channel $c_{\text{in}} = c_{\text{out}} = c= 64$, image feature map size $N = 256$.}
\label{tab:no_complexity}
\vspace{-1em}
\begin{tabular}{lccc}
\toprule
\textbf{Operator} & \textbf{Complexity per layer} & \textbf{FLOPs} & \textbf{Relative} \\
\midrule
FNO~\cite{fnop} & $\mathcal{O}(c^2 N^2 + c\,N^2 \log N)$ & $6.0 \times 10^{8}$ & $1\times$ \\
DISCO~\cite{liu2024neural} & $\mathcal{O}(c^2\, s\, N^2)$ & $2.4 \times 10^{9}$ & $4\times$ \\
LocalNO~\cite{liu2024neural} & $\mathcal{O}(c^2\, s\, N^2 + c^2 N^2 + c\,N^2 \log N)$ & $5.4 \times 10^{9}$ & $9\times$ \\
DeepONet~\cite{lu2019deeponet} & $\mathcal{O}\!\left(N^2(c\,d + L\,d^2 + d\,p\,c)\right)$ & $3.9 \times 10^{10}$ & $65\times$ \\
\bottomrule
\end{tabular}
\end{table}

\noindent \textbf{Why FNO is insufficient for imaging tasks.} Although FNO is the cheapest operator ($1\times$), it is equivalent to a global convolution in the spatial domain: every output pixel depends equally on all input pixels. As demonstrated in~\cite{liu2024neural}, this global mixing severely degrades performance on tasks that require spatially localized processing, such as image reconstruction, where preserving edges, textures, and fine anatomical structures is critical. Global operators over-smooth these local features, producing blurry reconstructions. Image reconstruction fundamentally requires a \emph{local integral operator}---one whose kernel has compact support so that each output is determined by a local neighborhood of the input. FNO lacks this property, making it unsuitable despite its low cost.

\noindent \textbf{DISCO is the most efficient neural local operator.} For tasks requiring local support, we compare DISCO against two alternatives. DeepONet~\cite{lu2019deeponet} lacks explicit local spatial structure and costs $65\times$ more than FNO, making it impractical for high-resolution imaging. LocalNO~\cite{liu2024neural}, the current state-of-the-art local neural operator, combines FNO with parallel differential and DISCO branches, achieving strong accuracy but at $9\times$ FNO's cost. Since DISCO is already the local component within LocalNO, it represents the minimal unit for introducing local inductive bias---at just $4\times$ FNO's cost, which is $2.25\times$ cheaper than LocalNO. Furthermore, as shown in Sec.~\ref{sec:runtimecnn}, DISCO achieves the same inference FLOPs as standard CNNs, since its continuous kernel is pre-discretized at test time and reduces to a regular convolution. This makes CTO competitive in efficiency with CNN-based methods while retaining the resolution-agnostic benefits of neural operators.

\subsection{Run Time Comparison with CNNs}
\label{sec:runtimecnn}

Since CTO employs only a single neural operator layer in each of $\NO_s$ and $\NO_i$, its computational cost closely matches that of a CNN with equivalent depth. A neural operator layer with DISCO convolution has the same asymptotic FLOP count as a standard convolutional layer with identical channel dimensions, as both scale as $\mathcal{O}(C_{\mathrm{in}} \cdot C_{\mathrm{out}} \cdot H \cdot W)$. The only additional cost arises during training from repeated kernel sampling; at inference, the discretized filter is pre-computed and fixed, making DISCO and CNN identical in runtime.

\noindent \textbf{Training.} During training, the continuous integration kernel must be sampled onto the discretized grid at each forward pass. At $256 \times 256$ resolution, a single DISCO layer requires $4.68$ MFLOPs compared to $3.28$ MFLOPs for the equivalent CNN layer---an increase of $42$\%. At $512 \times 512$, the corresponding figures are $63.45$ MFLOPs vs $52.63$ MFLOPs, a $20.5$\% increase. The relative overhead diminishes as spatial dimensions grow, since the kernel sampling cost is fixed while the convolution cost scales with the spatial size.

\noindent \textbf{Inference.} During inference, the discretized filter is pre-computed and fixed for the given input resolution---the continuous kernel does not need to be sampled again. The operation then reduces to a standard convolution, making DISCO and CNN identical in inference runtime. In Table~\ref{tab:complexity}, we compare GFLOPs/forward pass and parameter count across the Unrolled CNN baseline, a CTO variant with the identical architecture to the CNN ablation, and our original CTO model. At matched configuration, CTO and CNN have exactly the same FLOPs and parameter count, confirming that the continuous parameterization introduces no inference-time overhead. The full CTO model, whose kernel configuration is instead tuned for optimal performance, incurs a modest 1.3$\times$ FLOPs and 1.1$\times$ parameter overhead relative to the baseline, while improving 18-view PSNR from 29.14 to 31.57 dB.

\begin{table}[t]
\caption{Computational cost vs. accuracy. We compare GFLOPs per forward pass, parameter count, and 18-view PSNR for the Unrolled CNN baseline, a CTO variant using the identical architecture as the CNN ablation (\textit{Identical no. of param ablation}), and our full CTO model. At matched configuration, CTO and CNN have identical FLOPs and parameter count, confirming that the continuous parameterization introduces no inference-time overhead. Our full model instead tunes its kernel configuration for optimal performance, incurring a modest $1.3\times$ FLOPs and $1.1\times$ parameter overhead while improving PSNR by 2.43\,dB over the baseline.}
\label{tab:complexity}
\vspace{-1em}
\centering
\resizebox{1\textwidth}{!}{\begin{tabular}{lccc}
\toprule
Method & GFLOPS/forward pass & No. of parameters & 18-view PSNR \\
\midrule
Unrolled CNN & 36.33 (1$\times$) & 23.1\,M (1$\times$) &  29.14  \\
\method (Identical no. of param ablation) & 36.33 (1$\times$) & 23.1M (1$\times$) & 31.11 \\
\textbf{\method} & 47.53 (1.3$\times$) & 25.8\,M (1.1$\times$) & 31.57 \\
\bottomrule
\end{tabular}}
\vspace{-1em}
\end{table}

\end{document}